\documentclass[12pt,a4paper]{article}
\usepackage[T1]{fontenc}
\usepackage[sc,osf]{mathpazo}
\usepackage{a4wide}  
\usepackage{amsfonts}
\usepackage{amssymb}
\usepackage{amsmath}
\usepackage{bbm}
\usepackage{booktabs} 
\usepackage{ifpdf}
\ifpdf
\usepackage[pdftex,unicode,implicit]{hyperref}
\hypersetup{%
  pdftitle    = {N=2 Einstein-Yang-Mills' static two-center solutions}
  pdfkeywords = {gravity, supergravity, black hole, monopole, Wu-Yang, meron, 
multicenter, Yang-Mills,  Einstein-Yang-Mills, Higgs, non-Abelian, singular monopole, 
't~Hooft-Polyakov monopole, Bogomol'nyi},
  pdfauthor   = {Pablo Bueno, Pedro Fernandez-Ramirez, Patrick Meessen and Tomas Ortin},
  plainpages  = true,
  colorlinks  = true,
  citecolor   = blue,
  urlcolor    = red,
  linkcolor   = black
}
\newcommand{\hepth}[1]{{\tt
\href{http://www.arXiv.org/abs/hep-th/#1}{hep-th/#1}}}
\newcommand{\grqc}[1]{{\tt
\href{http://www.arXiv.org/abs/gr-qc/#1}{gr-qc/#1}}}

\newcommand{\arxiv}[1]{{\tt
\href{http://www.arXiv.org/abs/#1}{#1}}}
\else
  \usepackage[dvips]{graphicx}
  \usepackage[unicode,implicit]{hyperref}
  \newcommand{\hepth}[1]{{\tt hep-th/#1}}
  \newcommand{\grqc}[1]{{\tt gr-qc/#1}}
  
  \newcommand{\arxiv}[1]{{\tt arXiv:#1}}
\fi
\usepackage{tikz}\newcommand{\FPAUO}[2]{
\tikz[scale=.13,
         Uniovi/.style={color=green!51!blue, fill=green!51!blue}
 ] {
 \fill[Uniovi] (0,0) circle (10);
 \fill[white] (0,7) circle (1.5);
 \draw[Uniovi] (-2,7.5) rectangle (2,5.5);
 \fill[white] (-0.3,6.6) rectangle (0.3,0);   
 \fill[white] ( -0.9,6.2) rectangle (.9 ,5.6);
 \fill[white] (-1.4, 5.2) rectangle (1.4, 4.6);
 \fill[white] (0,0) ellipse (3.5 and 4);
 \fill[Uniovi] (-2.5,0.3) rectangle (2.5,-0.3);
 \fill[Uniovi] (-2,2.3) rectangle (2,1.7);
 \fill[Uniovi] (-2,-2.3) rectangle (2,-1.7);
 \fill[white] (-4.5,5.5) rectangle (-2.7,4.9);
 \fill[white] (-3.9,6.1) rectangle (-3.3,4.3);
 \fill[white] (4.5,5.5) rectangle (2.7,4.9);
 \fill[white] (3.9,6.1) rectangle (3.3,4.3);
 \foreach \x in { 0,..., 3 }
   \foreach \y in { 0,...,\x}
    {
     \fill[white] (-6-\x*0.7+\y*1.4,3.5-\x *1.97) -- (-5.6-\x*0.7+\y*1.4,2.4-\x *1.97) -- (-6.4-\x*0.7+\y*1.4,2.4-\x *1.97) -- cycle;
     \fill[white] (6-\x*0.7+\y*1.4,3.5-\x *1.97) -- (5.6-\x*0.7+\y*1.4,2.4-\x *1.97) -- (6.4-\x*0.7+\y*1.4,2.4-\x *1.97) -- cycle;
   };
 \draw (0,-6) node[
                               text centered, 
                               color=white, 
                               font={\fontsize{8}{4}\sffamily\selectfont}
                             ] {FPAUO-#1/#2};
}} 
\usepackage{pgfplots}    
\makeatletter
\@addtoreset{equation}{section}
\makeatother

\begin{document}

\begin{flushright}
\small
\FPAUO{14}{08}\\
IFT-UAM/CSIC-14-101\\
\texttt{arXiv:1410.4160 [hep-th]}\\
October 15\textsuperscript{th}, 2014\\
\normalsize
\end{flushright}

\vspace{1.5cm}

\begin{center}

{\Large {\bf $\mathcal{N}=2$ Einstein-Yang-Mills' static two-center solutions}}

\vspace{1.5cm}

\renewcommand{\thefootnote}{\alph{footnote}}
{\sl\large  Pablo Bueno$^{1}$}${}^{,}$\footnote{E-mail: {\tt p.bueno [at] csic.es}},
{\sl\large Patrick Meessen$^{2}$}${}^{,}$\footnote{E-mail: {\tt meessenpatrick [at] uniovi.es}},
{\sl\large Tom\'{a}s Ort\'{\i}n$^{1}$}${}^{,}$\footnote{E-mail: {\tt Tomas.Ortin [at] csic.es}}
{\sl\large and Pedro F.~Ram\'{\i}rez$^{1}$}${}^{,}$\footnote{E-mail: {\tt p.f.ramirez [at]  csic.es}},

\setcounter{footnote}{0}
\renewcommand{\thefootnote}{\arabic{footnote}}

\vspace{1.5cm}

${}^{1}${\it Instituto de F\'{\i}sica Te\'orica UAM/CSIC\\
C/ Nicol\'as Cabrera, 13--15,  C.U.~Cantoblanco, E-28049 Madrid, Spain}\\ \vspace{0.3cm}

${}^{2}${\it HEP Theory Group, Departamento de F\'{\i}sica, Universidad de Oviedo\\
  Avda.~Calvo Sotelo s/n, E-33007 Oviedo, Spain}\\

\vspace{1.8cm}


{\bf Abstract}

\end{center}

\begin{quotation}
  We construct \textit{bona fide} one- and two-center supersymmetric
  solutions to $\mathcal{N}=2$, $d=4$ supergravity coupled to SU$(2)$
  non-Abelian vector multiplets. The solutions describe black holes
  and global monopoles alone or in equilibrium with each other and
  exhibit non-Abelian hairs of different kinds. 
\end{quotation}

\newpage
\pagestyle{plain}

\tableofcontents

\newpage


\section*{Introduction}

Contrary to what one might think, multi-black hole solutions need not
be related to supersymmetry or, like in the case of Kastor and
Traschen's solution in Ref.~\cite{Kastor:1992nn}, fake-supersymmetry.
Proof of this is given by various solutions {\em e.g.} the ones
presented in Refs.~\cite{Bena:2009en} and \cite{Chimento:2013pka}.
The benefit of using supersymmetry, however, is that after a few
decades' worth of investigations there are workable recipes for
creating supersymmetric solutions, which greatly facilitates the
construction and study of multi-black hole solutions.

The construction is particularly straightforward in ungauged
$\mathcal{N}=2$, $d=4$ supergravity coupled to vector multiplets where
there are clear-cut rules for a supersymmetric multi-object solution
to give rise to a well-defined multi-black hole solution
\cite{Majumdar:1947eu,kn:Pa,Perjes:1971gv,Israel:1972vx,Hartle:1972ya,Denef:2000nb,Chrusciel:2005ve,Bellorin:2006xr}:
i) positive mass of the constituents, ii) the near-horizon limit has
to have definite entropy, iii) the 2$^{nd}$ law of thermodynamics must
hold in the coalescence of constituents, and iv) the Denef constraints
\cite{Denef:2000nb} must be satisfied.  Depending on the charges the
latter may constrain the distance between the constituents but it
always implies the absence of NUT charge.

The oft forgotten case of ungauged $\mathcal{N}=2$, $d=4$ supergravity
coupled to non-Abelian vector multiplets, which we will refer to as
$\mathcal{N}=2$ Einstein-Yang-Mills, is similar to the Abelian case in
that there is a well-defined recipe for constructing supersymmetric
solutions \cite{Huebscher:2007hj,Hubscher:2008yz}. However, the
construction of supersymmetric solutions is greatly hindered not only
by the fact that not every Abelian theory can be non-Abelianized, but
doubly more so by the fact that the supersymmetric recipe requires the
use of solutions of the (non-Abelian) Bogomol'nyi equation on
$\mathbb{R}^{3}$ \cite{Bogomolny:1975de}. Our lack of knowledge of the
space of all solutions to this equation is a serious limitation to the
application of the supersymmetric recipe: there exists a vast
literature on single monopole solutions, {\em i.e.\/} regular
single-center solutions to the Bogomol'nyi equation (see {\em e.g.\/}
Refs.~\cite{Sutcliffe:1997ec}). Depending on the chosen
$\mathcal{N}=2$, $d=4$ model, these can be used to construct globally
regular supergravity solutions known as \textit{global monopoles}. A
lot less is known about the singular solutions to the Bogomol'nyi
equation which are the ones which give rise to black holes with
different degrees of non-Abelian hair
\cite{Huebscher:2007hj,Hubscher:2008yz,Meessen:2008kb}. Finally, even
less is known about multi-center solutions to the Bogomol'nyi equation.
These are the ones we need in order to to apply the supersymmetric
recipe to the construction of multi-center supergravity solutions,
with centers that correspond to global monopoles or black holes.

Luckily enough, some explicit solutions are known.\footnote{Finite-energy,
  multi-center solutions of the Yang-Mills or Yang-Mills-Higgs system which do
  not satisfy the Bogomol'nyi equation like those in
  Refs.~\cite{Kleihaus:1999sx,Kleihaus:2003nj,Kleihaus:2003tn} are also
  known.} In this paper we are going to use the solutions of the
$\mathrm{SU}(2)$ Bogomol'nyi equation found by Cherkis and Durcan
\cite{Cherkis:2007jm} and Blair and Cherkis \cite{Blair:2010kz} (which we will
generalize by adding \textit{Protogenov hair} \cite{Meessen:2008kb}). These
solutions describe an 't Hooft-Polyakov (-Protogenov) monopole in the presence
of an arbitrary number of Dirac monopoles embedded in $\mathrm{SU}(2)$, all
having charge opposite to the one carried by the former. These solutions can
(in principle) give rise to supergravity solutions describing black holes in
the presence of a global monopole.  The construction of these solutions is, at
the same time, our main goal and our main result.

Before we start constructing multi-black hole solutions, however, it
is worth reviewing briefly some of the previous work on solutions of
YM theories coupled to gravity\footnote{For more comprehensive reviews
  see {\em e.g.\/} Refs.~\cite{Volkov:1998cc}.  }.  Most of the
previous work on this topic was focused on pure Einstein-Yang-Mills
(EYM) theories, (the minimal non-Abelian extension of the
Einstein-Maxwell theory), ignoring the possible existence of unbroken
supersymmetry which is, however, one of our main concerns here.

Soon after the discovery of the 't~Hooft-Polyakov monopole
\cite{'tHooft:1974qc,Polyakov:1974ek} several groups found solutions
to the pure EYM theory \cite{Yasskin:1975ag} whose $\mathrm{SU}(2)$
gauge field is that of the Wu--Yang $\mathrm{SU}(2)$ monopole
\cite{Wu:1967vp}. The metric of all these solutions is that of the
($dS$ or $AdS$) non-extremal Reissner-Nordstr\"om black hole and the
singularity in the gauge field (generically expected for static YM
solutions \cite{Deser:1976wq}) is covered by an event horizon.

This coincidence of the metrics is due to the relation between the
Wu--Yang $\mathrm{SU}(2)$ monopole and the non-Abelian embedding of
the Dirac monopole Eq.~(\ref{eq:WYmonopole}): they are related by a
singular gauge transformation and therefore give rise to exactly the
same energy-momentum tensor as it is gauge invariant whether the gauge
transformation is singular or not.  For this reason, these solutions
have been regarded as not truly non-Abelian, even though there are
potentially measurable differences, see {\em e.g.\/}
Refs.~\cite{Hasenfratz:1976gr,Canfora:2012ap}.

Finding less trivial (``genuinely or essentially non-Abelian'')
solutions proved much more difficult and a \textit{non-Abelian
  baldness theorem} stating that the only black-hole solutions of the
EYM $\mathrm{SU}(2)$ theory with a regular horizon and non-vanishing
magnetic charge had to be non-Abelian embeddings of the
Reissner--Nordstr\"om solution was proven in \cite{Galtsov:1989ip}.
This theorem was subsequently generalized to prove the absence of
regular monopole or dyon solutions to the EYM theory in
Refs.~\cite{Ershov:1991nv,Bizon:1992pi}.

An ``essentially non-Abelian'' solution, globally regular
\cite{Smoller:1991ag} to EYM theory had already been found: the
Bartnik-McKinnon particle \cite{Bartnik:1988am}.  The Bartnik-McKinnon
particle and its black hole-type generalizations \cite{Volkov:1989fi},
are in fact families of unstable solutions indexed by a discrete
parameter and evade the non-Abelian baldness theorem by being bald,
{\em i.e.\/} they have no asymptotic charge. It is worth pointing out
that even though these solutions are only known numerically, they have
been proven to exist \cite{Smoller:1993pe}.

The classification of the possible EYM solutions for the gauge group
$\mathrm{SU}(2)$ \cite{Smoller:1995nk} suggests that one has to add
more fields to the theory in order to get ``essentially non-Abelian''
black-hole or gravitating monopole solutions with non-vanishing
charges.  Investigations of solutions to the EYM theory coupled to a
Higgs field in the adjoint representation \cite{Lee:1991vy} in the
BPS-limit, a theory that is closer to the one we are going to study
than EYM, shows that a globally well-defined 't Hooft-Polyakov
monopole exists and furthermore the existence of other
Bartnik-McKinnon-like solutions.

As far as 4-dimensional supergravity is concerned we have the (supersymmetric)
Harvey-Liu \cite{Harvey:1991jr} and the Chamseddine-Volkov
\cite{Chamseddine:1997nm} regular gravitating monopole solutions to gauged
$\mathcal{N}=4$, $d=4$ supergravity; in $\mathcal{N}=2$, $d=4$ theories there are
analytical solutions describing global monopole solutions and non-Abelian
black hole solutions with and without asymptotic magnetic charge. Needless to
say, all the solutions mentioned in this little historical \textit{expos\'e}
describe the fields corresponding to a single object.  To our knowledge, there
are no known, essentially non-Abelian multi-object
analytic\footnote{Numerical, multi-center solutions have been found
  previously, though. See,
  \textit{e.g.}~Refs.~\cite{Kleihaus:2000hx,Kleihaus:2004wu}. Some of those
  solutions can be embedded in $\mathcal{N}=1,d=4$ supergravity. However,
  representing massive objects, they can never be supersymmetric in that
  theory. The embedding in higher-$\mathcal{N}$ supergravities is much more
  difficult (if possible at all).  We thank J.~Kunz for pointing these works
  to us.}  solutions and this article intends to fill this gap by constructing
static solutions describing the interplay between an 't Hooft-Polyakov
monopole and a Dirac monopole of opposite charge in two generic classes of
gauged $\mathcal{N}=2$, $d=4$ models.

It is convenient to stress that in the theories we have called
$\mathcal{N}=2$, $d=4$ SEYM the gauge group does not contain any part of the
R-symmetry group. Indeed, in general (ungauged) $\mathcal{N}=2$, $d=4$
theories, the global symmetry group G can be written as

\begin{equation}
\text{G}
=
\text{G}_{\text{V}} \times \text{G}_{\text{hyper}}
\times \text{SU}(2)_{\text{R}}\times \text{U}(1)_{\text{R}}\, ,
\end{equation}

\noindent
where $\text{G}_{\text{V}}$ and $\text{G}_{\text{hyper}}$ stand for
the isometry groups of the special and quaternionic K\"ahler manifolds
respectively. When a (necessarily non-Abelian) subgroup of
$\text{G}_{\text{V}}$ is gauged (as in $\mathcal{N}=2$, $d=4$ SEYM
theories) the scalar potential is positive semidefinite, which allows
for asymptotically De-Sitter and asymptotically flat solutions (such
as the ones we construct in this paper). This is in contradistinction
to theories in which a subgroup of $\mathrm{SU}(2)_{\rm R}$ (or the
complete $\mathrm{SU}(2)_{\rm R}$) is gauged via Fayet-Iliopoulos
terms\footnote{The overall $\mathrm{U}(1)_{R}$ group cannot be gauged
  in this way. The Abelian gaugings discussed in the literature deal
  with a subgroup $\mathrm{U}(1)\in \mathrm{SU}(2)_{\rm R}$.}  in
whose case the scalar potential becomes negative definite, the
solutions thus being asymptotically anti-De Sitter. Lately, an intense
effort has been devoted to the construction of black-hole solutions
of theories with Abelian gaugings (that is, theories in which a
subgroup $\mathrm{U}(1)\in \mathrm{SU}(2)_{\rm R}$ has been gauged);
see, for instance,
Refs.~\cite{Cacciatori:2009iz,Hristov:2010ri,Klemm:2012vm,Toldo:2012ec,Gnecchi:2013mja,Halmagyi:2014qza}
and references therein. The case in which the full
$\text{SU}(2)_{\text{R}}$ has been gauged remains as unexplored as
challenging, even though the general form of the timelike
supersymmetric solutions of this theory has been given in
Ref.~\cite{Meessen:2012sr}.

This paper is organized as follows: in Section~\ref{sec-N2SEYM} we
review the theories we are going to work with ($\mathcal{N}=2$, $d=4$
Super-Einstein-Yang-Mills theories) and the recipe for constructing
timelike supersymmetric solutions (black holes, in particular). In
Section~\ref{sec-single} we apply that recipe to construct single,
static supersymmetric black-hole and monopole solutions of two
particular examples of $\mathrm{SU}(2)$-gauged $\mathcal{N}=2$, $d=4$
SEYM: the $\overline{\mathbb{CP}}^{3}$ model (quadratic)
(\ref{sec-CP3} ) and the $\mathrm{ST}[2,4]$ model (cubic)
(\ref{page:ST2nmodels}). We use as seeds for these solutions the
single-center solutions of the Bogomol'nyi equations reviewed in
Section~\ref{sec-Protogenov}. In Section~\ref{sec-multibh} we
construct multi-black-hole solutions for the same models using the
multi-center solutions of the Bogomol'nyi equations reviewed in
Section~\ref{sec-multiprotogenov}. Our conclusions are contained in
Section~\ref{sec-conclusions}. In the Appendices we review a
particularly interesting single-center solution of the
$\mathrm{SU}(2)$ Bogomol'nyi equations which appears in different
guises: as a ``Lorentzian meron''
(Appendix~\ref{app-lorentzianmeron}), as the Wu-Yang monopole
(Appendix~\ref{app-WY}) or as a solution of the Skyrme model
(Appendix~\ref{app-skyrme}). A higher-charge generalization of this
solution is reviewed in Appendix~\ref{app-higherchargeWY}.


\section{$\mathcal{N}=2$, $d=4$ SEYM and its supersymmetric 
black-hole solutions  (SBHSs)}
\label{sec-N2SEYM}

In this section we are going to introduce the class of theories that
we have called $\mathcal{N}=2$, $d=4$ SEYM theories and we are going to
review the recipe to construct all their timelike supersymmetric
solutions, presented in Ref.~\cite{Huebscher:2007hj}. We shall be
extremely brief. The interested reader can find more details in
Refs.~\cite{Hubscher:2008yz,Freedman:2012zz,kn:PRC2}; our conventions
are those of Refs.~\cite{Huebscher:2007hj,Hubscher:2008yz,kn:PRC2}.


\subsection{The theory}
\label{sec-thetheory}

$\mathcal{N}=2$, $d=4$ SEYM theories can be seen as the simplest $\mathcal{N}=2$
supersymmetrization of the Einstein-Yang-Mills (EYM) theories. They are
nothing but theories of $\mathcal{N}=2$, $d=4$ supergravity coupled to $n$ vector
multiplets in which a (necessarily non-Abelian)\footnote{
  The theory becomes
  identical to the ungauged one when the gauge group is Abelian.
} 
subgroup of
the isometry group of the (Special K\"ahler) scalar manifold has been gauged
using some of the vector fields of the theory as gauge
fields\footnote{
  \label{foot:gauging} A global symmetry group can be gauged if
  it acts on the vector fields in the adjoint representation. Furthermore, it
  is required to be a symmetry of the prepotential; see {\em e.g.\/} ref.~\cite{Huebscher:2007hj}
  for more details.
}. 

We will only be concerned with the bosonic sector of the theory, which
consists on the metric $g_{\mu\nu}$, the vector fields
$A^{\Lambda}{}_{\mu}$ ($\Lambda=0,1,\cdots,n$) and the complex scalars
$Z^{i}$ ($i=1,\cdots,n$). The action of the bosonic sector reads

\begin{equation}
\label{eq:N2d4SEYMaction}
  \begin{array}{rcl}
  S[g_{\mu\nu}, A^{\Lambda}{}_{\mu}, Z^{i}] 
  & = & 
{\displaystyle\int} d^{4}x \sqrt{|g|}
  \left[R 
    +2\mathcal{G}_{ij^{*}}\mathfrak{D}_{\mu}Z^{i}\mathfrak{D}^{\mu}Z^{*\, j^{*}}
    +2\Im\mathfrak{m}\mathcal{N}_{\Lambda\Sigma} 
    F^{\Lambda\, \mu\nu}F^{\Sigma}{}_{\mu\nu}
  \right. \\
  & & \\
  & & \left. 
    -2\Re\mathfrak{e}\mathcal{N}_{\Lambda\Sigma}  
    F^{\Lambda\, \mu\nu}\star F^{\Sigma}{}_{\mu\nu}
    -V(Z,Z^{*})
  \right]\, .
\end{array}
\end{equation}

\noindent
In this expression, $\mathcal{G}_{ij^{*}}$ is the K\"ahler
metric, $\mathfrak{D}_{\mu}Z^{i}$ is the gauge-covariant derivative 

\begin{equation}
\mathfrak{D}_{\mu}Z^{i}
= 
\partial_{\mu}Z^{i}+gA^{\Lambda}{}_{\mu}k_{\Lambda}{}^{i}\, ,  
\end{equation}

\noindent
$F^{\Lambda}{}_{\mu\nu}$ is the vector field strength

\begin{equation}
F^{\Lambda}{}_{\mu\nu} 
= 
2\partial_{[\mu}A^{\Lambda}{}_{\nu]} 
-gf_{\Sigma\Gamma}{}^{\Lambda}A^{\Sigma}{}_{\mu}A^{\Gamma}{}_{\nu}\, ,
\end{equation}

\noindent
$\mathcal{N}_{\Lambda\Sigma}$ is the period matrix and, finally, $V(Z,Z^{*})$ is
the scalar potential

\begin{equation}
V(Z,Z^{*}) 
=
-{\textstyle\frac{1}{4}} g^{2}
\Im\mathfrak{m}\mathcal{N}^{\Lambda\Sigma}\mathcal{P}_{\Lambda}\mathcal{P}_{\Sigma}\, .
\end{equation}

Since the imaginary part of the period matrix is negative definite,
the scalar potential is positive semidefinite, which leads to
asymptotically-flat or -De Sitter solutions.

In the above equations, $k_{\Lambda}{}^{i}(Z)$ are the holomorphic
Killing vectors of the isometries that have been gauged\footnote{
  The employed notation associates a Killing vector to each value
  of the index $\Lambda$ in order to avoid the introduction of yet another
  class of indices and the embedding tensor (See {\em e.g.\/} the reviews \cite{Trigiante:2007ki}); 
  it is understood that not all the $k_{\Lambda}$ need to be non-vanishing.
}
and $\mathcal{P}_{\Lambda}(Z,Z^{*})$ the corresponding momentum maps,
which are related to the Killing vectors and to the K\"ahler potential
$\mathcal{K}$ by

\begin{eqnarray}
i\mathcal{P}_{\Lambda}
& = &
k_{\Lambda}{}^{i}\partial_{i}\mathcal{K} -\lambda_{\Lambda}\, ,   
\\
& & \nonumber \\
k_{\Lambda\, i^{*}} 
& = &
i\partial_{i^{*}}\mathcal{P}_{\Lambda}\, ,
\end{eqnarray}

\noindent
for some holomorphic functions $\lambda_{\Lambda}(Z)$.  Furthermore,
the holomorphic Killing vectors and the generators $T_{\Lambda}$ of
the gauge group satisfy the Lie algebras

\begin{equation}
[k_{\Lambda},k_{\Sigma}] = -f_{\Lambda\Sigma}{}^{\Gamma}k_{\Gamma}\, ,
\hspace{1cm}
[T_{\Lambda},T_{\Sigma}] = +f_{\Lambda\Sigma}{}^{\Gamma}T_{\Gamma}\, .
\end{equation}

For the gauge group $\mathrm{SU}(2)$, which is the only one we are
going to consider, we use lowercase indices\footnote{These will be a
  certain subset of those represented by $\Lambda, \Sigma,\ldots$.}
$a,b,c=1,2,3$ and the structure constants are
$f_{ab}{}^{c}=-\varepsilon_{abc}$, so

\begin{equation}
\label{eq:Lieantisu2}
[k_{a},k_{b}] = +\varepsilon_{abc}k_{c}\, ,
\hspace{1cm}
[T_{a},T_{b}] = -\varepsilon_{abc}T_{c}\, .
\end{equation}

We will use the fundamental representation, in which the generators are
proportional to the standard Pauli matrices\footnote{These are
\begin{equation}
\label{eq:Paulimatrices}
\sigma^{1} =
\left(
\begin{array}{cc}
0 & 1 \\
1 & 0 \\
\end{array}
\right)\!,
\hspace{.5cm}
\sigma^{2} =
\left(
\begin{array}{cc}
0 & -i \\
i & 0 \\
\end{array}
\right)\!,
\hspace{.5cm}
\sigma^{3} =
\left(
\begin{array}{cc}
1 & 0 \\
0 & -1 \\
\end{array}
\right)\, ,
\hspace{1cm}
   \sigma^{a}\sigma^{b} =\delta^{ab} +i \varepsilon^{abc}\sigma^{c}\, .
  \end{equation}
} 
 $\sigma^{a}$

\begin{equation}
\label{eq:su2generatorsfundamental}
T_{a}\equiv +\tfrac{i}{2} \sigma^{a}\, , 
\,\,\,\,
\Rightarrow
\,\,\,\,
\mathrm{Tr}(T_{a}T_{b}) = -\tfrac{1}{2}\delta_{ab}\, .
\end{equation}

The equations of motion of the theory can be written in the following form:

\begin{eqnarray}
\label{eq:Emnn2}
G_{\mu\nu}
+2\mathcal{G}_{ij^{*}}[\mathfrak{D}_{(\mu}Z^{i} \mathfrak{D}_{\nu )}Z^{*\, j^{*}}
-{\textstyle\frac{1}{2}}g_{\mu\nu}
\mathfrak{D}_{\rho}Z^{i}\mathfrak{D}^{\rho}Z^{*\, j^{*}}]
\nonumber \\
& & \nonumber \\
+4 \mathcal{M}_{MN}
\mathcal{F}^{M}{}_{\mu}{}^{\rho}\mathcal{F}^{N}{}_{\nu\rho}
+{\textstyle\frac{1}{2}}g_{\mu\nu}V(Z,Z^{*})
& = & 0,
\\
& & \nonumber \\
\label{eq:Ei2} 
\mathfrak{D}^{2}Z^{i} 
+\partial^{i}G_{\Lambda\, \mu\nu}\star F^{\Lambda\, \mu\nu}
+{\textstyle\frac{1}{2}} \partial^{i}V(Z,Z^{*})
& = & 
0,
\\
& & \nonumber \\
\label{eq:EmL}
\mathfrak{D}_{\nu} \star G_{\Lambda}{}^{\nu\mu}
+{\textstyle\frac{1}{4}}g
\left(
k_{\Lambda\, i^{*}}\mathfrak{D}_{\mu} Z^{*}{}^{i^{*}}+
k^{*}_{\Lambda\, i}\mathfrak{D}_{\mu} Z^{i} 
\right)
& = &
0\, ,
\end{eqnarray}

\noindent
where $G_{\Lambda\, \mu\nu}$ is the dual vector field strength

\begin{equation} 
G_{\Lambda}
\equiv 
\Re\mathfrak{e}\mathcal{N}_{\Lambda\Sigma} F^{\Sigma} 
+
\Im\mathfrak{m}\mathcal{N}_{\Lambda\Sigma}\, \star  F^{\Sigma}\,  ,
\end{equation}

\noindent
$\mathcal{F}^{M}{}_{\mu\nu}$ is the symplectic vector of vector field
strengths

\begin{equation}
\left(\mathcal{F}^{M}\right) 
\equiv 
\left(
\begin{array}{c}
F^{\Lambda} \\ G_{\Lambda} \\ 
\end{array}
\right)\, ,
\end{equation}

\noindent
$\mathcal{M}_{MN}$ is the symmetric $2(n+1)\times 2(n+1)$ matrix
defined  by

\begin{equation}
(\mathcal{M}_{MN})
\equiv 
\left(
  \begin{array}{cc}
\Im\mathfrak{m}\mathcal{N}_{\Lambda\Sigma} 
+R_{\Lambda\Gamma}\Im\mathfrak{m}\mathcal{N}^{-1|\Gamma\Omega}R_{\Omega\Sigma}\,\,\,\,
& 
-R_{\Lambda\Gamma}\Im\mathfrak{m}\mathcal{N}^{-1|\Gamma\Sigma} 
\\
& \\
-\Im\mathfrak{m}\mathcal{N}^{-1|\Lambda\Omega}R_{\Omega\Sigma} 
& 
\Im\mathfrak{m}\mathcal{N}^{-1|\Lambda\Sigma}
\\
\end{array}
\right)\, ,
\end{equation}

\noindent
and 

\begin{equation}
\mathfrak{D}_{\nu} \star G_{\Lambda}{}^{\nu\mu}  
=
\partial_{\nu} \star G_{\Lambda}{}^{\nu\mu}
+
gf_{\Lambda\Sigma}{}^{\Gamma}A^{\Sigma}{}_{\nu}\star G_{\Lambda}{}^{\nu\mu}\, .
\end{equation}

Most of the literature and earlier work on non-Abelian black-hole and
monopole solutions has been carried out in the context of the
Einstein-Yang-Mills (EYM) and Einstein-Yang-Mills-Higgs (EYMH)
theories. Before closing this introduction, it is worth discussing the
relation between those and the theories we are considering here.
The main differences of the latter w.r.t.~the former are the complexification of the Higgs field 
and the presence
of a non-trivial period matrix. A further difference is the
possibility of having more general scalar manifolds, which is
reflected in the expressions of the gauge-covariant derivatives of the
scalar fields.  Solutions to the $\mathcal{N}=2$, $d=4$ SEYM theory have
a chance of being also solutions of the EYMH theory if they have
covariantly-constant scalars with identical phases (\textit{e.g.}~all
of them purely imaginary). Then, if the scalar potential vanishes on
the solutions, they also have a chance of being solutions to the EYM
system as well; as we are going to see, some of the solutions found in
Refs.~\cite{Huebscher:2007hj,Hubscher:2008yz} are also solutions of
the EYM theory and have the same metric as
the EYM solutions of
Refs.~\cite{Yasskin:1975ag,Canfora:2012ap}.

\subsection{The recipe to construct SBHSs of  $\mathcal{N}=2$, $d=4$ SEYM}
\label{sec-recipe}

To construct timelike supersymmetric solutions of the
$\mathcal{N}=2$, $d=4$ SEYM theory, it suffices to follow this recipe
\cite{Huebscher:2007hj,Hubscher:2008yz} to find the elementary
building blocks of the solutions, which are the $2(n+1)$
time-independent functions
$(\mathcal{I}^{M}) =\left(
  \begin{smallmatrix}
    \mathcal{I}^{\Lambda} \\ \mathcal{I}_{\Lambda} \\
  \end{smallmatrix}
\right)
$:

\begin{enumerate}
\item Take a solution of the Bogomol'nyi equations

\begin{equation}
\label{eq:Bogomolnyin2d4-bis}
\tilde{F}^{\Lambda}{}_{\underline{m}\underline{n}} 
= 
-\tfrac{1}{\sqrt{2}}\varepsilon_{mnp} 
\tilde{\mathfrak{D}}_{\underline{p}}\mathcal{I}^{\Lambda},
\end{equation}

\noindent
for a gauge field $\tilde{A}^{\Lambda}{}_{\underline{m}}$
($\underline{m}=1,2,3$ labels the 3 spatial coordinates) and a real
``Higgs'' field $\mathcal{I}^{\Lambda}$.
$\tilde{\mathfrak{D}}_{\underline{p}}\mathcal{I}^{\Lambda}$ is the
covariant derivative in the adjoint representation with gauge field
$\tilde{A}^{\Lambda}{}_{\underline{m}}$.  Observe that this equation
has to be solved in the gauged (non-Abelian) and ungauged (Abelian)
directions. The integrability condition in the Abelian directions is
the familiar requirement that the $\mathcal{I}^{\Lambda}$ be harmonic
functions on $\mathbb{R}^{3}$.

\item Find the functions $\mathcal{I}_{\Lambda}$
  by solving these equations:

\begin{equation}
\label{eq:DDILd2}
\tilde{\mathfrak{D}}_{\underline{m}}
\tilde{\mathfrak{D}}_{\underline{m}} \mathcal{I}_{\Lambda} 
=
\tfrac{1}{2}
g^{2}
\left[
f_{\Lambda (\Sigma}{}^{\Gamma}f_{\Delta )\Gamma}{}^{\Omega}\ 
\mathcal{I}^{\Sigma}\mathcal{I}^{\Delta}
\right]\, \mathcal{I}_{\Omega}\, .
\end{equation}

In the non-Abelian directions these equations can, in many cases, be solved by taking
$\mathcal{I}_{\Lambda} \propto \mathcal{I}^{\Lambda}$, but currently we only know how
to generate non-trivial solutions to them in the cases where the gauge doublet
$(\tilde{A}^{\Lambda}, \mathcal{I}^{\Lambda})$ describes a non-Abelian Wu-Yang monopole;
Observe that $\mathcal{I}_{\Lambda}=0$ is always a solution, but the physical
fields may be singular in some models.

In the Abelian directions, the $\mathcal{I}_{\Lambda}$ are just
independent harmonic functions on $\mathbb{R}^{3}$.

\item Given the functions $\mathcal{I}^{M}$, we must find the 1-form
  on $\mathbb{R}^{3}$ $\omega_{\underline{m}}$ by solving the
  following equation:

\begin{equation}
\label{eq:omega}
\partial_{[\underline{m}}\omega_{\underline{n}]} 
=
\varepsilon_{mnp}
\mathcal{I}_{M} \tilde{\mathfrak{D}}_{\underline{p}}\mathcal{I}^{M}
=
\varepsilon_{mnp}
\left(
\mathcal{I}_{\Lambda} \tilde{\mathfrak{D}}_{\underline{p}}\mathcal{I}^{\Lambda}
-
\mathcal{I}^{\Lambda} \tilde{\mathfrak{D}}_{\underline{p}}\mathcal{I}_{\Lambda}
\right)\, .
\end{equation}

The integrability conditions of this equation impose constraints on
the integration constants of the functions
$\mathcal{I}^{M}$ in exactly the same manner as in the ungauged case
\cite{Denef:2000nb,Bates:2003vx}. 

In the case of static solutions, {\em i.e.\/} when $\omega=0$, the above equation becomes a
constraint on the integration constants of the functions
$\mathcal{I}^{M}$ that will have to be solved. Observe, however, that this constraint is
independent of the specific $\mathcal{N}=2$, $d=4$ model and only depends on the
choice of gauge group; possible restrictions on the solution to said constraint can come
from the desired behaviour of the physical fields in the full solution.

\item To reconstruct the physical fields from the functions
  $\mathcal{I}^{M}$ we need to solve the stabilization equations,
  a.k.a.~\textit{Freudenthal duality equations}, which give the
  components of the Freudenthal dual\footnote{
    In
    Refs.~\cite{Meessen:2006tu,Huebscher:2007hj,Hubscher:2008yz} the
    components of the Freudenthal dual are denoted by
    $\mathcal{R}^{M}$.
}  
$\tilde{\mathcal{I}}^{M}(\mathcal{I})$ in terms of the functions $\mathcal{I}^{M}$ \cite{Galli:2012ji};
These relations completely characterize the model of $\mathcal{N}=2$, $d=4$ supergravity.
\par
Equivalently, the $\tilde{\mathcal{I}}$ can be derived from a homogeneous function of degree 2
$W(\mathcal{I})$ called the {\em Hesse potential} as \cite{Bates:2003vx,Mohaupt:2011aa,Meessen:2011aa}
\begin{equation}
  \label{eq:HessePotential}
  \tilde{\mathcal{I}}_{M}\; =\; \textstyle{1\over 2}\frac{\partial W}{\partial \mathcal{I}^{M}} \;\;\;\longrightarrow\;\;\;
  W(\mathcal{I})\; =\; \tilde{\mathcal{I}}_{M}\mathcal{I}^{M} \; .
\end{equation}

\item The metric takes the form 

\begin{equation}
ds^{2}
=
e^{2U} (dt+\omega)^{2} -e^{-2U}dx^{m}dx^{m}\, ,
\end{equation}

\noindent
where $\omega=\omega_{\underline{m}}dx^{m}$ is the above spatial 1-form
and the metric function $e^{-2U}$ is given by

\begin{equation}
e^{-2U} 
=  
\tilde{\mathcal{I}}_{M}(\mathcal{I}) \mathcal{I}^{M}
= W(\mathcal{I}) \, .
\end{equation}

\item The scalar fields are given by

\begin{equation}
Z^{i}
=
\frac{\tilde{\mathcal{I}}^{i}+i\mathcal{I}^{i}}{\tilde{\mathcal{I}}^{0}+i\mathcal{I}^{0}}\, .    
\end{equation}

\item The components of the vector fields are given by 

\begin{eqnarray}
A^{\Lambda}{}_{t}
& = &
-\tfrac{1}{\sqrt{2}}e^{2U}\tilde{\mathcal{I}}^{\Lambda}\, ,
\\
& & \nonumber \\
A^{\Lambda}{}_{\underline{m}}
& = &
\tilde{A}^{\Lambda}{}_{\underline{m}} +\omega_{\underline{m}}\ A^{\Lambda}{}_{t}\, .
\end{eqnarray}



\end{enumerate}

After having gone through the steps of the recipe, one ends up with a supersymmetric solution to 
a chosen $\mathcal{N}=2$, $d=4$ EYM theory and what remains to be done is to analyze the constraints
coming from imposing appropriate regularity conditions such as the absence of naked singularities.

\section{Static, single-SBHSs of $\mathrm{SU}(2)$ $\mathcal{N}=2$, $d=4$ 
SEYM and pure EYM}
\label{sec-single}

Following the recipe given in Section~\ref{sec-recipe}, we are going
to construct static, single-center SBHSs of $\mathrm{SU}(2)$
$\mathcal{N}=2$, $d=4$ SEYM. Some of the solutions will simultaneously
solve the equations of motion of the EYM and EYMH theories.

The first step consists in finding a solution
$\tilde{A}^{\Lambda}{}_{\underline{m}},\mathcal{I}^{\Lambda}$ of the
$\mathrm{SU}(2)$ Bogomol'nyi equations in $\mathbb{R}^{3}$
Eqs.~(\ref{eq:Bogomolnyin2d4-bis}).


\subsection{Single-center solutions of the $\mathrm{SU}(2)$ 
Bogomol'nyi equations in $\mathbb{R}^{3}$}
\label{sec-Protogenov}

Before we search for solutions of the Bogomol'nyi equations it is
worth reviewing the origin and meaning of those equations in the
context of the $\mathrm{SU}(2)$ Yang-Mills-Higgs theory (in the
Bogomol'nyi-Prasad-Sommerfield (BPS) limit in which the Higgs
potential vanishes).


\subsubsection{The $\mathrm{SU}(2)$ Yang-Mills-Higgs system}

With the normalization in Eq.~(\ref{eq:su2generatorsfundamental}) and
writing $F \equiv F^{a}T_{a}, \Phi \equiv \Phi^{a}T_{a}$, the action
of the YMH theory in our conventions reads

\begin{equation}
\label{eq:YMHaction}
S_{\rm YMH} 
=
-2\int  d^{4}x\, 
\mathrm{Tr}\,\left\{ \tfrac{1}{2} \mathfrak{D}_{\mu}\Phi
  \mathfrak{D}^{\mu}\Phi 
-\tfrac{1}{4}F_{\mu\nu}F^{\mu\nu}\right\}\, ,
\end{equation}

\noindent
and the corresponding equations of motion are

\begin{eqnarray}
\mathfrak{D}_{\mu} F^{\mu\nu} 
& = & 
g[\Phi, \mathfrak{D}^{\nu}\Phi]\, ,
\\
& & \nonumber \\
\mathfrak{D}^{2} \Phi
& = & 
0\, .  
\end{eqnarray}

For static configurations $F_{t\underline{m}} = \mathfrak{D}_{t}\Phi=0$,
the action can be written, up to a total derivative, in the manifestly
positive form

\begin{equation}
S_{\rm YMH} 
=
-2\int  d^{4}x\, 
\mathrm{Tr}\,\left\{
-\tfrac{1}{4}
\left( F_{\underline{m}\underline{n}} \mp \varepsilon_{mnp}\mathfrak{D}_{\underline{p}}\Phi \right)
\left( F_{\underline{m}\underline{n}} \mp \varepsilon_{mnp}\mathfrak{D}_{\underline{p}}\Phi \right) 
\right\}\, ,  
\end{equation}

\noindent
which leads to the conclusion that static field configurations satisfying the
first-order \textit{Bogomol'nyi equations} \cite{Bogomolny:1975de}

\begin{equation}
\label{eq:Beqs}
F_{\underline{m}\underline{n}} =\pm \varepsilon_{mnp}\mathfrak{D}_{\underline{p}}\Phi\, ,
\end{equation}

\noindent
extremize the action Eq.~(\ref{eq:YMHaction}) and are solutions of the
full Yang-Mills-Higgs equations. Indeed, if we act with
$\mathfrak{D}_{\underline{m}}$ on both sides of the equation and use
the Ricci identity and the Bogomol'nyi equation we get the Yang-Mills
equation:

\begin{equation}
\mathfrak{D}_{\underline{m}}F_{\underline{m}\underline{n}} 
=
\mp \varepsilon_{nmp}\mathfrak{D}_{\underline{m}}\mathfrak{D}_{\underline{p}}\Phi
= 
\mp
\tfrac{1}{2}g\varepsilon_{nmp} [F_{\underline{m}\underline{p}},\Phi]
= 
-g[\mathfrak{D}_{\underline{n}}\Phi,\Phi]\, . 
\end{equation}

\noindent
If, instead, we act with
$\varepsilon_{pmn}\mathfrak{D}_{\underline{p}}$ and use the Bianchi
identity, we get the Higgs equation:

\begin{equation}
0
=
\varepsilon_{pmn}\mathfrak{D}_{\underline{p}}F_{\underline{m}\underline{n}} 
=
\pm \mathfrak{D}_{\underline{p}} \mathfrak{D}_{\underline{p}}\Phi\, .
\end{equation}

Observe that the source of the Yang-Mills field, the \textit{Higgs current}
$g[\Phi, \mathfrak{D}\Phi]$, not only vanishes when the Higgs
field is covariantly constant $\mathfrak{D}\Phi=0$ but also when $\Phi$ and
$\mathfrak{D}\Phi$ are parallel in $\mathfrak{su}(2)$.

Eqs.~(\ref{eq:Beqs}) are identical to the ones that arise in
$\mathcal{N}=2$, $d=4$ SEYM theory, (\ref{eq:Bogomolnyin2d4-bis}) upon
the identification of the vector fields and 

\begin{equation}
\label{eq:identif}
\tfrac{1}{\sqrt{2}}\mathcal{I}^{a} = \mp \Phi^{a}\, .  
\end{equation}


\subsubsection{The hedgehog ansatz}

In order to construct static, single-center black-hole-type
solutions, it is natural to look for spherically symmetric solutions
of Eqs.~(\ref{eq:Beqs}). Substituting the \textit{hedgehog ansatz}

\begin{equation}
\mp \Phi^{a} = \delta^{a}{}_{m}f(r)x^{m}\, ,
\hspace{1cm}
A^{a}{}_{\underline{m}} = -\varepsilon^{a}{}_{mn}x^{n} h(r)\,   
\end{equation}

\noindent
in the Bogomol'nyi Eqs.~(\ref{eq:Beqs})  we get an equivalent
system of differential equations for $f(r)$ and $h(r)$:

\begin{equation}
\begin{array}{rcl}
r\partial_{r}h +2h  - f(1+gr^{2}h) 
& = & 
0\, ,
\\
& & \\
r\partial_{r}(h+f) -gr^{2}h (h+ f)
& = & 
0\, .
\\
\end{array}
\end{equation}

After Prasad and Sommerfield \cite{Prasad:1975kr} found the solution describing the 't Hooft-Polyakov
monopole in the BPS limit, Protogenov \cite{Protogenov:1977tq} classified all spherically symmetric solutions
to the $\mathrm{SU}(2)$ Bogomol'nyi equations: the ones that can be used to generate
BH-like spacetimes are a 2-parameter family
$(f_{\mu,s},h_{\mu,s})$ plus a 1-parameter family
$(f_{\lambda},h_{\lambda})$ given by

\begin{equation}
\label{eq:Protogenovsolutions}
\begin{array}{rclrcl}
rf_{\mu,s} 
& = & 
{\displaystyle
\frac{1}{gr}
\left[
1-\mu r\coth{(\mu r+s)} 
\right]
}\, ,
\hspace{1cm}
&
rh_{\mu,s} 
& = &
{\displaystyle 
\frac{1}{gr}
\left[
\frac{\mu r}{\sinh{(\mu r+s)}} 
-1
\right]
}\, ,
\\
& & & & & \\
rf_{\lambda} 
& = & 
{\displaystyle
\frac{1}{gr}
\left[
\frac{1}{1+\lambda^{2}r}
\right]\, ,
}
&
rh_{\lambda}
& = & 
-rf_{\lambda}\, .
\end{array}
\end{equation}

\noindent
The parameter $s$ is known in the black-hole context as the
\textit{Protogenov hair} parameter \cite{Meessen:2008kb}. The BPS
't~Hooft-Polyakov monopole \cite{Prasad:1975kr} is the only globally regular solution of
this family (which explains why it is the only one usually considered
in the monopole literature\footnote{
  After coupling the system to
  gravity, the singularities of the other solutions may become
  ``harmless'' if they can be covered by regular event horizons.
})  
and corresponds to $s=0$.  In the $s \rightarrow \infty$ limit we get

\begin{equation}
\label{eq:sinftysolution}
-rf_{\mu,\infty} 
=  
\frac{\mu}{g} -\frac{1}{gr}, 
\hspace{1cm}
rh_{\mu,\infty} 
=
-\frac{1}{gr}\, ,
\end{equation}

\noindent
which, for $\mu=0$, coincides with the Wu-Yang monopole
\cite{Wu:1967vp} given in Eq.~(\ref{eq:WYmonopole}), and is a
solution of the pure Yang-Mills theory. This is possible because the
Higgs current $g[\Phi, \mathfrak{D}\Phi]$ vanishes even though $\Phi$
is neither zero nor covariantly constant\footnote{Actually,
  the only field configuration in this ansatz with a vanishing Higgs
  current is this one.}. 
With a non-trivial Higgs field, though, we
can assign a well-defined monopole charge to it: for any $\mu$ and $s$

\begin{equation}
\frac{1}{4\pi}\int_{S^{2}_{\infty}}\mathrm{Tr}(\hat{\Phi} F)=\frac{1}{g}\, ,
\hspace{1cm}
\hat{\Phi} \equiv \frac{\Phi}{\sqrt{|\mathrm{Tr}(\Phi^{2})|}}\, .  
\end{equation}

The same field configuration can be seen as a Lorentzian meron (see
Appendix~\ref{app-lorentzianmeron}) and as a solution to the Skyrme
model (see Appendix~\ref{app-skyrme}), and, crucially, it is related
to the $\mathrm{SU}(2)$-embedded Dirac monopole by a singular gauge
transformation (see Appendix~\ref{app-WY}).  Since the
metric is oblivious to gauge transformations, singular or not, the
Wu-Yang monopole gives rise to solutions whose metric is identical to
that of Abelian case.\footnote{
 Of course there are measurable differences between these two situations, see
 {\em e.g.\/} Refs.~\cite{Hasenfratz:1976gr,Canfora:2012ap}.
}
The same applies to the higher-charge
generalizations of the Lorentzian meron/Wu-Yang monopole reviewed in
Appendix~\ref{app-higherchargeWY}.

If fact, this mechanism can be used to generate Wu-Yang monopoles of
higher charge from the well-known Dirac monopole solutions of charge
higher than 1 embedded in $\mathrm{SU}(2)$, as reviewed in
Appendix~\ref{app-higherchargeWY}. The metric cannot see the
difference between the non-Abelian and the Abelian fields given in Eq.~(\ref{eq:sinftysolution}).

The 1-parameter family is singular for all values of the parameter
$\lambda$, which also appears in black-hole solutions as hair. The
magnetic charge measured at spatial infinity vanishes according to the
above definition. However, it can be argued that these solutions do
describe a magnetic monopole placed at the origin whose charge is
screened: the entropy of black hole associated to this field has the
same form as that of the black hole associated to the Wu-Yang
monopole. Observe that, for $\lambda=0$, the solution is identical to
the Wu-Yang monopole with $\mu=0$, Eqs.~(\ref{eq:sinftysolution}).


\subsubsection{The Protogenov trick}
\label{sec:PTrick}

As it turns out, many regular monopole solutions can be deformed by adding a
parameter $s$ to the argument $\mu r$, generating a family of
solutions that contains the original one ($s=0$) and, typically, a new
and simpler solution in the $s\rightarrow \infty$ limit. We will refer
to this procedure as \textit{the Protogenov trick} and it can be
justified as follows: let us consider, for instance, the
't~Hooft-Polyakov monopole.  Since the Bogomol'nyi equation is
polynomial, having elementary functions such as hyperbolic functions
in the solution means that they must cancel amongst themselves and
that only their derivatives contribute to the polynomial part of the
solution. This means that one should be able to deform the dependency
of the elementary functions introducing a shift $s$ of the radial
coordinate and still solve the Bogomol'nyi equations.

Of course, the cancellations necessary for having a regular solution
will not work out anymore (assuming they did work for $s=0$) and one
will end up with a family of singular solutions. We will use this
trick later.


\subsection{Embedding in the  $\mathrm{SU}(2)$-gauged $\overline{\mathbb{CP}}^{3}$ model}
\label{sec-CP3} 


\subsubsection{The $\overline{\mathbb{CP}}^{3}$ model}

The $\overline{\mathbb{CP}}^{n}$ models have $n$ vector supermultiplets and
are defined by the quadratic prepotentials

\begin{equation}
\mathcal{F} 
=
-\tfrac{i}{4}\eta_{\Lambda\Sigma}
\mathcal{X}^{\Lambda}\mathcal{X}^{\Sigma}\, ,
\hspace{1cm}
(\eta_{\Lambda\Sigma}) = \mathrm{diag}(+-\dotsm -)\, .
\end{equation}

\noindent
The $n$ physical scalar fields can be defined as

\begin{equation}
Z^{i} \equiv \mathcal{X}^{i}/\mathcal{X}^{0}\, ,
\end{equation}

\noindent
and they parametrize the symmetric space
$\mathrm{U}(1,n)/(\mathrm{U}(1)\times \mathrm{U}(n))$.  It is convenient to define
$Z^{0}\equiv 1$, $Z^{\Lambda}\equiv
\mathcal{X}^{\Lambda}/\mathcal{X}^{0}$ and $Z_{\Lambda}\equiv
\eta_{\Lambda\Sigma}Z^{\Sigma}$.  In the $\mathcal{X}^{0}=1$ gauge,
the K\"ahler potential and the K\"ahler metric are given by

\begin{equation}
\label{eq:Kcpn}
\mathcal{K} 
 =  
-\log{(Z^{*\Lambda}Z_{\Lambda})}\, ,
\hspace{.4cm}
\mathcal{G}_{ij^{*}} 
= 
-e^{\mathcal{K}}\left( \eta_{ij^{*}}
-e^{\mathcal{K}}Z^{*}_{i}Z_{j^{*}} \right)\, ,
\,\,\,\,
\Rightarrow
\,\,\,\,
0\leq \sum_{i}|Z^{i}|^{2}< 1\, .  
\end{equation}

\noindent
The above metric is the standard (Bergman) metric for the
U$(1,n)/($U$(1)\times$U$(n))$ symmetric spaces \cite{Besse:1987pua}.
The covariantly holomorphic symplectic section $\mathcal{V}$ and the
period matrix $\mathcal{N}_{\Lambda\Sigma}$ are given by

\begin{equation}
\label{eq:Vcpn}
\mathcal{V}
=
e^{\mathcal{K}/2}
\left(
  \begin{array}{c}
  Z^{\Lambda} \\ \\ -\tfrac{i}{2} Z_{\Lambda} \\
  \end{array}
\right)\, ,
\hspace{1cm}
\mathcal{N}_{\Lambda\Sigma}
=
\tfrac{i}{2}
\left[
\eta_{\Lambda\Sigma}
-
2\frac{Z_{\Lambda}Z_{\Sigma}}{Z^{\Gamma}Z_{\Gamma}}
\right]\, . 
\end{equation}

The isometry subgroup $\mathrm{SU}(1,n)$ acts linearly, in the fundamental
representation, on the coordinates $\mathcal{X}^{\Lambda}$

\begin{equation}
\mathcal{X}^{\prime\,\Lambda}
= 
\Lambda^{\Lambda}{}_{\Sigma}\mathcal{X}^{\Sigma}\, ,   
\hspace{.5cm}
\mathrm{with}
\hspace{.5cm}
\Lambda^{\dagger}\eta \Lambda=\eta\, ,
\hspace{.5cm}
\mathrm{and}
\hspace{.5cm}
\det \Lambda=1\, .
\end{equation}

\noindent 
This linear action induces a non-linear action on the special coordinates:

\begin{equation}
\label{refsca}
Z^{ \prime\, \Lambda}
=
\frac{\Lambda^{\Lambda}{}_{\Sigma}Z^{\Sigma}}{\Lambda^{0}{}_{\Sigma}
Z^{\Sigma}}\, .
\end{equation}

\noindent
The  K\"ahler potential is invariant under these transformations up to
K\"ahler transformations $\mathcal{K}^{\prime}=\mathcal{K}+f+f^{*}$ with

\begin{equation}
f(Z)
=
\log{\left(\Lambda^{0}{}_{\Sigma} Z^{\Sigma} \right)}\, .  
\end{equation}

The $n(n+2)$ infinitesimal generators $T_{m}$ of $\mathfrak{su}(1,n)$ in the
fundamental representation are defined by

\begin{equation}
\label{eq:infinitesimalSU1n}
\Lambda^{\Lambda}{}_{\Sigma}
\sim
\delta^{\Lambda}{}_{\Sigma}+\alpha^{m}\ T_{m}{}^{\Lambda}{}_{\Sigma}\, ,
\hspace{.5cm}
\mathrm{with}
\hspace{.5cm}
\eta T^{\dagger}_{m}\eta =-T_{m}\, ,
\hspace{.5cm}
\mathrm{and}
\hspace{.5cm}
T_{m}{}^{\Lambda}{}_{\Lambda}
=
0\, .
\end{equation}

\noindent
Substituting this definition into Eq.~(\ref{refsca}) we find an expression
for the holomorphic Killing vectors\footnote{The $k_{m}{}^{0}(Z)$ component
  vanishes identically, as it must, but it is convenient to keep it.}.

\begin{equation}
Z^{\prime\, \Lambda}
=
Z^{\Lambda}+\alpha^{m} k_{m}{}^{\Lambda}(Z)\, ,
\hspace{1cm}
k_{m}{}^{\Lambda}(Z)
=
T_{m}{}^{\Lambda}{}_{\Sigma}\ Z^{\Sigma}
- T_{m}{}^{0}{}_{\Omega}\  Z^{\Omega} Z^{\Lambda}\, ,   
\end{equation}

\noindent
and, from this expression, we also find explicit expressions for the
holomorphic functions $\lambda_{m}(Z)$ and the momentum maps

\begin{equation}
\label{eq:lmPmCPn}
\lambda_{m} 
= 
T_{m}{}^{0}{}_{\Sigma}Z^{\Sigma}\, ,
\hspace{1cm}
\mathcal{P}_{m} 
=
ie^{\mathcal{K}}
T_{m}{}^{\Lambda}{}_{\Sigma}Z^{\Sigma}Z^{*}_{\Lambda}
=
ie^{\mathcal{K}}
\eta_{\Lambda\Omega}T_{m}{}^{\Lambda}{}_{\Sigma}Z^{\Sigma}Z^{*\,
  \Omega}\, .    
\end{equation}

Although the theory is invariant under the whole SU$(1,n)$ group, the
prepotential is invariant only under the subgroup of
$\mathrm{SU}(1,n)$ with real matrices, $\mathrm{SO}(1,n)$, which is
the largest group that we could eventually gauge. However, the
requirements that the vectors must transform in the adjoint
representation restricts the possibilities to either $\mathrm{SO}(1,2)$
or $\mathrm{SO}(3)$ (if $n\geq 2$ or $n\geq 3$, respectively); 
we are going to consider the latter case embedded into the minimal
model admitting this gauge group, namely 
$\overline{\mathbb{CP}}^{3}$.

In this model, the adjoint indices $a,b,c,\ldots$ and the fundamental indices
$i,j,k,\ldots$ take the same values $1,2,3$ and we will only use the
latter. The infinitesimal transformations of the scalars are

\begin{equation}
\delta_{\alpha}Z^{i} 
= 
\alpha^{j}T_{j}{}^{i}{}_{k} Z^{k}\, ,   
\,\,\,\,\,
\mbox{where}
T_{j}{}^{i}{}_{k} = f_{jk}{}^{i} = -\epsilon_{jki}\, ,
\end{equation}

\noindent
and the momentum maps, holomorphic Killing vectors etc. take the values

\begin{equation}
\mathcal{P}_{i} 
=
-ie^{\mathcal{K}}\epsilon_{ijk}Z^{j}Z^{*\, k}\, ,
\hspace{1cm}
k_{i}{}^{j} 
= 
\epsilon_{ijk}Z^{k}\, ,
\hspace{1cm}
\lambda_{i}
=0\, .
\end{equation}

\noindent
This means that the gauge-covariant derivative of the scalars is just that
of a complex adjoint $\mathrm{SO}(3)$ scalar

\begin{equation}
\mathfrak{D}_{\mu}Z^{i} 
= 
\partial_{\mu}Z^{i} -g\epsilon_{ijk}A^{j}{}_{\mu}Z^{k},  
\end{equation}

\noindent
and that the scalar potential takes the form

\begin{equation}
  \label{eq:scalarpotentialCP3}
  V(Z,Z^{*}) \; =\;
        -\tfrac{1}{2}g^{2} e^{\mathcal{K}} 
          \epsilon_{ijk}\epsilon_{imn}Z^{j}Z^{* k^{*}}Z^{m}Z^{* n^{*}} 
     \; =\;
      \tfrac{1}{2}g^{2}\ \left| \vec{Z}\times \vec{Z}^{*}\right|^{2} \; .   
\end{equation}


\subsubsection{The solutions}

To construct the solutions of this model\footnote{All these solutions
  have already been presented in
  Refs.~\cite{Huebscher:2007hj,Meessen:2008kb,Hubscher:2008yz}. We
  review them here for pedagogical reasons and also for the sake of
  making easier the comparison with the solutions of other models.} we
just have to follow the recipe spelled out in
Section~\ref{sec-recipe}. We will only consider static solutions (so
$\omega=0$ and $\tilde{A}^{\Lambda}{}_{\underline{m}} =
A^{\Lambda}{}_{\underline{m}}$). First of all, we need a solution of
the Bogomol'nyi Eqs.~(\ref{eq:Bogomolnyin2d4-bis}). These equations
split into an Abelian part (the $0$th component) and the non-Abelian
part (the $i=1,2,3$ components):

\begin{eqnarray}
F^{0}{}_{\underline{m}\underline{n}} 
& =  &
-\tfrac{1}{\sqrt{2}}\epsilon_{mnp} 
\partial_{\underline{p}}\mathcal{I}^{0}\, ,
\\
& & \nonumber \\
F^{i}{}_{\underline{m}\underline{n}}
& = & 
-\tfrac{1}{\sqrt{2}}\epsilon_{mnp} 
\mathfrak{D}_{\underline{p}}\mathcal{I}^{i}\, .
\end{eqnarray}

\noindent
The Abelian equation is solved by 

\begin{equation}
\mathcal{I}^{0}= A^{0}+\frac{p^{0}/\sqrt{2}}{r}\, ,  
\end{equation}

\noindent
where $A^{0}$ is an integration constant and $p^{0}$ is the normalized Abelian
magnetic charge. The non-Abelian set of equations can be solved making the
identification Eq.~(\ref{eq:identif}) and using  Protogenov's solutions
Eqs.~(\ref{eq:Protogenovsolutions}).

The second step in the recipe (finding solutions $\mathcal{I}_{\Lambda}$ to
Eqs.~(\ref{eq:DDILd2})) will be solved, for the sake of simplicity, by choosing
another harmonic function in the Abelian direction and vanishing functions in
the rest:

\begin{equation}
\mathcal{I}_{0}= A_{0}+\frac{q_{0}/\sqrt{2}}{r}\, ,  
\hspace{1cm}
\mathcal{I}_{i}=0\, .
\end{equation}

The third point in the recipe, combined with the staticity of the solutions
implies the following constraint on the integration constants:

\begin{equation}
\label{eq:omega=0nonA}
A^{0}q_{0}-A_{0}p^{0}=0\, .  
\end{equation}

\noindent
Another constraint will arise from the normalization of the metric at
infinity. The solution is completely determined and, now, we only have to
write the physical fields and make, if necessary, sensible choices of the
values of the physical parameters to make the solutions regular.

In order to write the physical fields we need the solutions of the Freudenthal
duality equations of this model. These are given by (see,
\textit{e.g.}~Ref.~\cite{Bueno:2013pja})

\begin{equation}
\label{eq:FdualCPn}
(\tilde{\mathcal{I}}^{M}) 
=
\left(
  \begin{array}{c}
\tilde{\mathcal{I}}^{\Lambda} \\ 
\tilde{\mathcal{I}}_{\Lambda} \\
  \end{array}
\right)
=
\left(
  \begin{array}{c}
-2\eta^{\Lambda\Sigma}\mathcal{I}_{\Sigma} \\ 
\tfrac{1}{2}\eta_{\Lambda\Sigma} \mathcal{I}^{\Sigma} \\
  \end{array}
\right)\, ,
\,\,\,\,\,
\Rightarrow
\,\,\,\,\,
e^{-2U}
=
\tfrac{1}{2}\eta_{\Lambda\Sigma}\mathcal{I}^{\Lambda}\mathcal{I}^{\Sigma}
+2\eta^{\Lambda\Sigma}\mathcal{I}_{\Lambda}\mathcal{I}_{\Sigma}\, ,
\end{equation}

\noindent
and the metric function and the physical scalars are given by

\begin{eqnarray}
e^{-2U} 
& = & 
\tfrac{1}{2}(\mathcal{I}^{0})^{2} +2(\mathcal{I}_{0})^{2} -(rf)^{2},
\\
& & \nonumber \\
\label{eq:scalarsnonAbeliansolutions}
Z^{i} 
& = &
\frac{\sqrt{2} rf}{\mathcal{I}^{0}+2i\mathcal{I}_{0}} \delta^{i}{}_{m}\frac{x^{m}}{r}. 
\end{eqnarray}

At least one of the two functions $\mathcal{I}^{0},\mathcal{I}_{0}$ must be
different from zero for the metric function to be positive. Then, there are
two possible cases, depending on the vanishing of the Abelian charges
$p^{0},q_{0}$:

\begin{description}
\item[I. $p^{0}=q_{0}=0$] The only regular solution is the one with $s=0$ (the
  't~Hooft-Polyakov monopole). In this solution, the integration constants
  satisfy the normalization condition

\begin{equation}
\tfrac{1}{2}(A^{0})^{2}+2(A_{0})^{2} = 1 +(\mu/g)^{2}\, . 
\end{equation}

\noindent
They are also related to the asymptotic values of the scalars. These
cannot be constant, in general, because the scalars transform under
local $\mathrm{SU}(2)$ transformations, but they are covariantly
constant and their asymptotic values are determined by a single
gauge-invariant complex parameter that we call
$Z_{\infty}$:\footnote{Observe that the scalar potential of this
  theory, Eq.~(\ref{eq:scalarpotentialCP3}), vanishes at infinity for
  those solutions, which is why they are asymptotically flat.}

\begin{equation}
Z^{i} 
\sim 
Z_{\infty} \delta^{i}{}_{m}\frac{x^{m}}{r}\ ,
\hspace{1cm}
Z_{\infty} 
\equiv 
\frac{\mu/g}{ 1 +(\mu/g)^{2}}\left(\tfrac{1}{\sqrt{2}}A^{0}-\sqrt{2}
  iA_{0} \right)\, ,
\hspace{1cm}  
0\leq |Z_{\infty}|^{2}< 1\, .
\end{equation}

These expressions lead to the following identification of the integration
constant $\mu$ in terms of the physical parameters:

\begin{equation}
\label{eq:muinCP3}
\mu^{2} = \frac{|Z_{\infty}|^{2}}{1-|Z_{\infty}|^{2}}g^{2}\, ,   
\end{equation}

\noindent
and to the following expression for the mass of the solution

\begin{equation}
\label{eq:massmonopole}
M_{\rm monopole}
= 
\sqrt{\frac{|Z_{\infty}|^{2}}{1-|Z_{\infty}|^{2}}}\frac{1}{g}\, .
\end{equation}

This asymptotically flat solution has no singularities nor horizons (one finds
a Minkowski spacetime in the $r\rightarrow 0$ limit, hence zero entropy and
temperature). Globally-regular solutions of this kind
\cite{Harvey:1991jr,Chamseddine:1997nm} are sometimes
called \textit{global monopoles}.

Observe that a solution of the ungauged theory with 

\begin{equation}
H^{0}=A^{0}\, ,
\hspace{.5cm}
H_{0}=A_{0}\, ,
\hspace{.5cm}
H^{1}=A^{1}+\frac{\sqrt{2}}{gr}\, ,
\end{equation}

\noindent
in which the non-Abelian monopole is replaced by an Abelian monopole
with the same charge, would have the same asymptotic behavior but it
would mean having a naked singularity at some value of $r>0$.

\item[II. $p^{0}q_{0}\neq 0$]\footnote{It is easier to work with both
    charges non-vanishing. The results will still be valid when we set
    one of them to zero.} Solving Eq.~(\ref{eq:omega=0nonA}) the
  metric can be written in the form

\begin{eqnarray}
e^{-2U} 
& = & 
\frac{1}{1-|Z_{\infty}|^{2}}H^{2} -(rf)^{2},
\\
& & \nonumber \\
Z^{i} 
& = &
\frac{2\beta}{p^{0}+2iq_{0}}
\frac{rf}{H} \delta^{i}{}_{m}\frac{x^{m}}{r}\, , 
\end{eqnarray}

\noindent
where $H$ is the harmonic function 

\begin{equation}
H\equiv 1+\frac{\beta}{r}\, ,
\hspace{1cm}
\beta^{2}
=
(1-|Z_{\infty}|^{2})W_{\rm RN}(\mathcal{Q})/2\, ,
\hspace{1cm}
 W_{\rm RN}(\mathcal{Q})
\equiv 
\tfrac{1}{2}(p^{0})^{2} +2(q_{0})^{2}\, ,
\end{equation}

\noindent
and the integration constant $\mu$ is still given by
Eq.~(\ref{eq:muinCP3}). We have expressed all the constants (except for
Protogenov's hair parameter $s$ and $\lambda$) in terms of physical constants.
Observe that the isolated solution $f_{*}$ has $\mu=0$ and corresponds to
$Z_{\infty}=0$. These identifications allow us to compute the mass and entropy
of all the possible solutions in terms of the physical parameters. We get a
completely general mass formula and two formulae for the entropy, one for the
$s\neq 0$ solutions and another one for the $s=0$ and the isolated solutions
(which corresponds to $Z_{\infty}=0$):

\begin{eqnarray}
M
& = & 
\sqrt{\tfrac{1}{2}\frac{W_{RN}(\mathcal{Q})}{1-|Z_{\infty}|^{2}}}
+
M_{\rm monopole},
\\
& & \nonumber \\
S/\pi
& = &
\tfrac{1}{2}
W_{\rm RN}(\mathcal{Q})
-\frac{1}{g^{2}}\, ,
\hspace{.5cm}
\mbox{for}
\hspace{.5cm}
s\neq 0
\,\,\,\,
\mbox{and}
\,\,\,\,
Z_{\infty}=0,
\\
& & \nonumber \\
S/\pi
& = &
\tfrac{1}{2}
W_{\rm RN}(\mathcal{Q}),
\hspace{.5cm}
\mbox{for}
\hspace{.5cm}
s=0\, , 
\end{eqnarray}

\noindent
where $M_{\rm monopole}$ is given by Eq.~(\ref{eq:massmonopole}). 

The entropy is moduli-independent as in the ungauged case and both the entropy
and the mass are independent of the hair parameters $s$ and $\lambda$.

Observe that the charge of the BPS 't~Hooft-Polyakov monopole $s=0$
does not contribute to the entropy which suggests that it must be
associated to a pure state in the quantum theory. The non-Abelian
field of the isolated solution does not contribute to the mass at
infinity ($M_{\rm monopole}=0$ for $Z_{\infty}=0$) but there is a
magnetic-charge contribution to the entropy, which suggests that there
really is a magnetic charge but it is screened at infinity. Further
support for this interpretation comes from the near-horizon limit of
the scalars, which is the covariantly-constant function of the charges

\begin{equation}
Z^{i}_{\rm h} 
= 
\frac{1/g}{\tfrac{1}{2}p^{0}+iq_{0}}\ \delta^{i}{}_{m}\frac{x^{m}}{r}.
\end{equation}

\noindent
even for the isolated case, when no magnetic charge is observed at infinity.

In the case of the 1-parameter ($\lambda$) family, neither the mass
nor the entropy depend on $\lambda$.

\end{description}

Some of the solutions in this family can also be seen as solutions of
the pure EYM theory. They are identical to those obtained in
Refs.~\cite{Yasskin:1975ag,Canfora:2012ap}. As
discussed at the end of Section~\ref{sec-thetheory}, we need to tune
the parameters of the solutions so as to get covariantly constant
scalars which do not contribute to the energy-momentum tensor This is
only possible for the $s\rightarrow \infty$ solutions (Wu--Yang
monopoles) for which $rf$ is a harmonic function. In that case

\begin{equation}
Z^{i} 
= 
Z\ \delta^{i}{}_{m}\frac{x^{m}}{r}\, ,
\hspace{1cm}
Z
= 
\frac{1/g}{\tfrac{1}{2}p^{0}+i q_{0}}=Z_{\infty}\, .  
\end{equation}

The metric is identical to that of a Reissner-Nordstr\"om black hole.
These solutions were called \textit{black hedgehogs} in
Ref.~\cite{Huebscher:2007hj} and \textit{black merons} in
Ref.~\cite{Canfora:2012ap} because the gauge field of the Wu--Yang monopole
can also be understood as Lorentzian meron solution. 

A closely related solution with non-covariantly constant scalars was obtained
in a different context in Ref.~\cite{Kallosh:1994wy}.


\subsection{Embedding in $\mathrm{SU}(2)$-gauged $\mathrm{ST}[2,n]$ models}


\subsubsection{The $ST[2,n]$ models}
\label{page:ST2nmodels}

The $ST[2,n]$ models are cubic models with $n_{V}=n+1$ vector supermultiplets
and as many complex scalars and, as all other cubic models, they can be
embedded in type~II String Theory compactified Calabi-Yau 3-folds and then
uplifted to M-theory. They can also be obtained from corresponding models of
$N=1$, $d=5$ supergravity compactified on $S^{1}$.

A generic cubic model is defined by the prepotential
\begin{equation}
  \label{eq:2}\displaystyle
  \mathcal{F} \; =\; -\frac{1}{3\text{!}}\ d_{ijk}\ \frac{\mathcal{X}^{i}\mathcal{X}^{j}\mathcal{X}^{k}}{\mathcal{X}^{0}} \; ,
\end{equation}
where $d$ is completely symmetric in its indices;
the $ST[2,n]$ models are characterized by $d$-tensors with non-vanishing
components $d_{1\alpha\beta}= \eta_{\alpha\beta}$ where $(\eta_{\alpha\beta})
= \mathrm{diag}(+-\dotsm -)$ and where the indices $\alpha,\beta$ take $n$
values between $2$ and $n+1$.

The scalar $Z^{1}=\mathcal{X}^{1}/\mathcal{X}^{0}$ plays a special role and
parametrizes a SL$(2,\mathbb{R})/$SO$(2)$ coset space. For this and other
reasons, it is called axidilaton and we will denote it by $\tau$. The other
$n$ scalars parametrize a SO$(2,n)/($SO$(2)\times$SO$(n))$ coset space and
will be denoted by $Z^{\alpha}=\mathcal{X}^{\alpha}/\mathcal{X}^{0}$
($\alpha=2,\cdots,n$).  The K\"ahler metric and 1-form
connection are the products of those of the two spaces.

Using this notation and using the gauge $\mathcal{X}^{0}=1$, the canonical
symplectic section $\Omega$, the K\"ahler potential $\mathcal{K}$ and the
components of K\"ahler 1-form $\mathcal{Q}_{i}$ and of the K\"ahler metric
$\mathcal{G}_{ij^{*}}$ are given by

\begin{equation}
\begin{array}{rclrcl}
\Omega 
& = &
\left(
  \begin{array}{c}
 1 \\ 
\tau \\ 
Z^{\alpha} \\ 
\tfrac{1}{2}\tau\eta_{\alpha\beta}Z^{\alpha}Z^{\beta} \\ 
-\tfrac{1}{2}\eta_{\alpha\beta}Z^{\alpha}Z^{\beta} \\ 
-\tau\eta_{\alpha\beta}Z^{\beta} \\ 
  \end{array}
\right),
\hspace{.5cm}
&
e^{-\mathcal{K}} 
& = & 
4\Im \mathfrak{m}\, \tau\ \eta_{\alpha\beta}\Im \mathfrak{m}\, Z^{\alpha}\ \Im
\mathfrak{m}\, Z^{\beta},
\\
& & & & & \\
\mathcal{Q}_{\tau} 
& = &
{\displaystyle
\frac{1}{4\Im\mathfrak{m}\, \tau}
}, 
&
\mathcal{Q}_{\alpha} 
& = &
{\displaystyle
\frac{\eta_{\alpha\beta} \Im
\mathfrak{m}\, Z^{\beta}}{2\eta_{\gamma\delta}\Im \mathfrak{m}\, Z^{\gamma}\ 
\Im\mathfrak{m}\, Z^{\delta}}\, ,
}
\\
& & & & & \\
\mathcal{G}_{\tau\tau^{*}} 
& = & 
{\displaystyle
\frac{1}{4(\Im\mathfrak{m}\, \tau)^{2}}
},
&
\mathcal{G}_{\alpha\beta^{*}} 
& = & 
{\displaystyle
\frac{\eta_{\alpha\gamma} \Im\mathfrak{m}\, Z^{\gamma}\ 
\eta_{\beta\delta} \Im\mathfrak{m}\,
Z^{\delta}}{\left[\eta_{\epsilon\varphi}\Im \mathfrak{m}\, Z^{\epsilon}\
  \Im\mathfrak{m}\, Z^{\varphi}\right]^{2}} -\frac{\eta_{\alpha\beta}}{2 \eta_{\epsilon\varphi}\Im \mathfrak{m}\, Z^{\epsilon}\
  \Im\mathfrak{m}\, Z^{\varphi}}\, .
}
\end{array}
\end{equation}

The reality of the K\"ahler potential constrains the values of the
scalars. The model has two branches characterized by

\begin{equation}
\Im \mathfrak{m}\, \tau >0\, ,
\hspace{1cm}
 \eta_{\alpha\beta}\Im \mathfrak{m}\, Z^{\alpha}\ \Im\mathfrak{m}\, Z^{\beta}>0\, ,
\end{equation}

\noindent
and

\begin{equation}
\Im \mathfrak{m}\, \tau\ <0\, ,
\hspace{1cm}
\eta_{\alpha\beta}\Im \mathfrak{m}\, Z^{\alpha}\ \Im\mathfrak{m}\, Z^{\beta}<0\, ,
\end{equation}

\noindent
that will be distinguished where required by $+$ and $-$ indices,
respectively.

Only the subgroup SO$(1,n)\subset\, $SO$(2,n)$ acts linearly (in the
fundamental representation) on the special coordinates $Z^{\alpha}$ and the
group SO$(3)$ acts in the adjoint (for instance) on the coordinates
$\alpha=3,4,5$ if $n\geq 4$. We take $n=4$ for simplicity and denote the
$\alpha=3,4,5$ indices by $a,b,\cdots=1,2,3$. For the sake of simplicity we
will write $Z^{a}$ instead of $Z^{a+2}$ for $Z^{3},Z^{4},Z^{5}$ etc.  The
generators and structure constants of $\mathfrak{so}(3)$ and their action on
the scalars are the same as in the $\overline{\mathbb{CP}}^{3}$ model with
obvious changes of notation:

\begin{equation}
(T_{a})^{b}{}_{c} = f_{ac}{}^{b} = -\varepsilon_{acb}\, ,  
\hspace{1cm}
\delta_{\alpha}Z^{a} = \alpha^{b}(T_{b})^{a}{}_{c}Z^{c}  
=
-\epsilon_{abc}\alpha^{b}Z^{c} = \alpha^{b} k_{b}{}^{a}(Z)\, ,
\end{equation}

\noindent
($\tau$ and $Z^{2}$ are inert) so the holomorphic Killing vectors and the
momentum maps are

\begin{equation}
k_{a}{}^{b}(Z) 
= 
\epsilon_{abc}Z^{c}\, ,
\hspace{1cm}
\mathcal{P}_{a}
=
-\tfrac{i}{2}
\frac{\epsilon_{abc}Z^{b}Z^{*\, c^{*}}}{\eta_{\alpha\beta}
\Im \mathfrak{m}\, Z^{\alpha}\ \Im\mathfrak{m}\, Z^{\beta}}\, .
\end{equation}

The scalar potential has a structure similar to that of the
$\overline{\mathbb{CP}}^{3}$ model, but more complicated. We will not give it
here since it is not needed anyway.


\subsubsection{The solutions}

To find solutions to this non-Abelian model we just need to follow the
recipe. First, we find the functions $\mathcal{I}^{\Lambda}$ and the spatial
components of the vector fields $A^{\Lambda}{}_{\underline{m}}$ by solving the
Bogomol'nyi equations

\begin{eqnarray}
F^{\Lambda}{}_{\underline{m}\underline{n}} 
& =  &
-\tfrac{1}{\sqrt{2}}\epsilon_{mnp} 
\partial_{\underline{p}}\mathcal{I}^{\Lambda}\, ,
\hspace{.5cm}
I=0,1,2,
\\
& & \nonumber \\
F^{a+2}{}_{\underline{m}\underline{n}}
& = & 
-\tfrac{1}{\sqrt{2}}\epsilon_{mnp} 
\mathfrak{D}_{\underline{p}}\mathcal{I}^{a+2}\, ,
\hspace{.5cm}
a=1,2,3,
\end{eqnarray}

\noindent
(we will suppress the $+2$ in the non-Abelian indices in most places).  The
Abelian equations are solved by harmonic functions and the non-Abelian ones by
making the identification Eq.~(\ref{eq:identif}) with the Higgs field and
using Protogenov's solutions Eqs.~(\ref{eq:Protogenovsolutions}), as we did in
the $\overline{\mathbb{CP}}^{3}$ model. 

Next, we have to find the functions $\mathcal{I}_{\Lambda}$ by solving
Eqs.~(\ref{eq:DDILd2}). In the Abelian directions $\Lambda=0,1,2$ we can
simply choose harmonic functions and in the non-Abelian ones we take
$\mathcal{I}_{a}=0$. This choice gives non-singular solutions, as we are going
to see.  We will also set some of the harmonic functions to zero for
simplicity.

The Hesse potential defined in Eq.~(\ref{eq:HessePotential}) can be found from 
Shmakova's solution of the stabilization
(or Freudenthal duality) equations for cubic models \cite{Shmakova:1996nz}; it
can be written as

\begin{equation}
\mathsf{W}(\mathcal{I})
=
2 \sqrt{J_{4}(\mathcal{I})}\, ,
\end{equation}

\noindent
with the quartic invariant $J_{4}(\mathcal{I})$ given by 

\begin{equation}
J_{4}(\mathcal{I})
\equiv
(\mathcal{I}^{\alpha}\mathcal{I}^{\beta}\eta_{\alpha\beta}
+2\mathcal{I}^{0}\mathcal{I}_{1})
(\mathcal{I}_{\alpha}\mathcal{I}_{\beta}\eta^{\alpha\beta}
-2\mathcal{I}^{1}\mathcal{I}_{0})
-(\mathcal{I}^{0}\mathcal{I}_{0}-\mathcal{I}^{1}\mathcal{I}_{1}
+\mathcal{I}^{\alpha}\mathcal{I}_{\alpha})^{2}\, .
\end{equation}

This potential does not vanish for the choice $\mathcal{I}_{a}=0$, as we
advanced and it will remain non-singular if we set
$\mathcal{I}^{0}=\mathcal{I}_{1}=\mathcal{I}_{2}=0$. In other words: the only
non-trivial components of $\mathcal{I}^{M}$ are
$\mathcal{I}^{1},\mathcal{I}^{2},\mathcal{I}^{a+2},\mathcal{I}_{0}$. With this
choice the metric function is given by

\begin{equation}
\label{metric}
e^{-2U}
=
\mathsf{W}(\mathcal{I})
=
2\sqrt{-2\mathcal{I}^1\mathcal{I}_{0}\,
\eta_{\alpha\beta}\mathcal{I}^{\alpha}\mathcal{I}^{\beta}}
=
2\sqrt{-2\mathcal{I}^1\mathcal{I}_{0}
[(\mathcal{I}^{2})^{2}-\mathcal{I}^{a}\mathcal{I}^{a}]
}\, .
\end{equation}



As instructed by the recipe in Sec.~(\ref{sec-recipe}), we can calculate the $\tilde{\mathcal{I}}$ from Eq.~(\ref{eq:HessePotential}), which 
for our choice of non-trivial components of $\mathcal{I}^{M}$ means that
$\tilde{\mathcal{I}}^{i}=0$ ($i=1,\cdots,5$); this implies that all the scalars are purely
imaginary and given by

\begin{equation}
Z^{i}
=
i\frac{\mathcal{I}^{i}}{\tilde{\mathcal{I}}^{0}}\, ,
\hspace{.5cm}
\mbox{where}
\hspace{.5cm}
\tilde{\mathcal{I}}^{0}
=
\frac{2\mathcal{I}^{1}
\eta_{\alpha\beta}\mathcal{I}^{\alpha}\mathcal{I}^{\beta}}{\mathsf{W}(\mathcal{I})}\, .
\end{equation}

\noindent
It is convenient to write all of them in terms of $\tau=Z^{1}$

\begin{equation}
\label{scalars}
Z^{\alpha} 
=
\frac{\mathcal{I}^{\alpha}}{\mathcal{I}^{1}} \tau\, ,
\hspace{1cm}
\tau
=
i\frac{e^{-2U}}{2\eta_{\alpha\beta}\mathcal{I}^{\alpha}\mathcal{I}^{\beta}}\, .   
\end{equation}

In the two ($+$ and $-$) branches of the model
corresponding, respectively, to the upper and lower signs
$\pm\Im\mathfrak{m}\, \tau_{(\pm)} >0$ and, since $e^{-2U}>0$, we must
choose the functions $\mathcal{I}^{\alpha}_{(\pm)}$ so
that

\begin{equation}
\pm \eta_{\alpha\beta}\mathcal{I}^{\alpha}_{(\pm)}\mathcal{I}^{\beta}_{(\pm)} 
=
\pm \left[(\mathcal{I}_{(\pm)}^{2})^{2}
-\mathcal{I}_{(\pm)}^{a}\mathcal{I}^{a}_{(\pm)}\right]>0\, .  
\end{equation}

\noindent
{}In order for $\mathsf{W}(\mathcal{I})$ to be real the $\mathcal{I}_{(\pm)\, 0}$ and
$\mathcal{I}_{(\pm)}^{1}$ must be chosen so as to satisfy

\begin{equation}
\pm \mathcal{I}^{1}_{(\pm)}\mathcal{I}_{(\pm)\,   0} < 0\, .
\end{equation}

(We will suppress the $\pm$ subindices in what follows, to simplify the
notation, except where this may lead to confusion.)

Observe that with our choice of non-vanishing components of $\mathcal{I}^{M}$
the r.h.s.~of Eq.~(\ref{eq:omega}) vanishes automatically, whence the staticity condition
$\omega =0$ does not impose any constraint.

According to the preceding discussions, the non-vanishing components of
$\mathcal{I}^{M}$ will be assumed to take the form 

\begin{equation}
\begin{array}{rclrclrcl}
\mathcal{I}^{1} & = & A^{1}+{\displaystyle\frac{p^{1}/\sqrt{2}}{r}}\, , \hspace{.6cm}&
\mathcal{I}^{2} & = & A^{2}+{\displaystyle\frac{p^{2}/\sqrt{2}}{r}}\,  ,\hspace{.6cm}&
\mathcal{I}^{a} & = & \sqrt{2}\, \delta^{a}{}_{m}x^{m}f(r)\, ,\\
& & & & & & & & \\
\mathcal{I}_{0} & = & A_{0}+{\displaystyle\frac{q_{0}/\sqrt{2}}{r}}\, ,
& & & & & & \\
\end{array}
\end{equation}

\noindent
where $f(r)$ is $f_{\mu,s}$ or $f_{\lambda}$ in
Eqs.~(\ref{eq:Protogenovsolutions}), $p^{1},p^{2},q_{0}$ are magnetic
and electric charges and $A^{1},A^{2},A_{0}$ are integration constants
to be determined in terms of the asymptotic values of the scalars and
the metric. These constants must have the same sign as the
corresponding charges 

\begin{equation}
\mathrm{sign}(A^{1,2}) = \mathrm{sign}(p^{1,2})\, ,
\hspace{1cm}
\mathrm{sign}(A_{0})= \mathrm{sign}(q_{0})\, ,
\end{equation}

\noindent
as the functions $\mathcal{I}^{1}$, $\mathcal{I}^{2}$ and $\mathcal{I}_{0}$
are required to have no zeroes on the interval $r\in (0,+\infty)$ in order to avoid naked singularities there. 
Then, the above
constraint on the signs of $\mathcal{I}^{1}$ and $\mathcal{I}_{0}$
translates into the following constraints on the signs of the charges
in the two branches:

\begin{equation}
\mathrm{sign}(p^{1})\mathrm{sign}(q_{0})= \mp 1\, .  
\end{equation}

Defining as in the $\overline{\mathbb{CP}}^{3}$ case the asymptotic
value $Z_{\infty}$ of the adjoint scalars by

\begin{equation}
Z^{a}_{\infty} \equiv Z_{\infty}\ \delta^{a}{}_{m}\frac{x^{m}}{r}\, ,
\end{equation}

\noindent
and imposing the normalization of the metric at infinity it is not
hard to express the integration constants $\mu,A^{1},A^{2},A_{0}$ in
terms of the moduli (the asymptotic values of the scalars
$\Im\mathfrak{m}\tau_{\infty},\Im\mathfrak{m}\, Z^{2}_{\infty}$ and
$\Im\mathfrak{m}\, Z_{\infty}$) and the coupling constant $g$

\begin{equation}
\begin{array}{rcl}
A^{1}
& = & 
{\displaystyle
\frac{\mathrm{sign}(p^{1})|\Im\mathfrak{m}\tau_{\infty}|}{\sqrt{2}\chi_{\infty}}    
}\, ,
\\
& & \\
A^{2}
& = & 
{\displaystyle
\frac{\mathrm{sign}(p^{2})|\Im\mathfrak{m}\, Z^{2}_{\infty}|}{\sqrt{2}\chi_{\infty}}    
}\, ,
\\
& & \\
\mu
& = & 
{\displaystyle
\frac{g|\Im\mathfrak{m}\, Z_{\infty}|}{2\chi_{\infty}}    
}\, ,
\\
& & \\
A_{0}
& = & 
\tfrac{1}{2\sqrt{2}}
\mathrm{sign}(q_{0})\chi_{\infty}\, ,
\\
& & \\
\end{array}
\end{equation}

\noindent
where we have defined the combination (real in both branches of the theory)

\begin{equation}
\chi_{\infty}  
\equiv
\sqrt{\Im\mathfrak{m}\tau_{\infty}\left[(\Im\mathfrak{m}\, Z^{2}_{\infty})^{2}
-(\Im\mathfrak{m}\, Z_{\infty})^{2} \right]}\, .
\end{equation}

\noindent
The mass of the solutions in terms of the moduli and the charges is

\begin{equation}
  \begin{array}{rcl}
M  
& = &
\tfrac{1}{4}
{\displaystyle\frac{\chi_{\infty}}{|\Im\mathfrak{m}\tau_{\infty}|}}
|p^{1}|
+
{\displaystyle\frac{1}{2\chi_{\infty}}|q_{0}|}
\pm\tfrac{1}{2} 
{\displaystyle
\frac{|\Im\mathfrak{m}\tau_{\infty}\Im\mathfrak{m}\, Z^{2}_{\infty}|}{\chi_{\infty}}
}|p^{2}|
\pm
{\displaystyle
\frac{|\Im\mathfrak{m}\tau_{\infty}\Im\mathfrak{m}\, Z_{\infty}|}{\chi_{\infty}}
\frac{1}{g}
}\, .
\end{array}
\end{equation}

\noindent
In the above expressions  we have used two consistency conditions:

\begin{equation}
\mathrm{sign}(\Im\mathfrak{m}\, Z_{\infty})  
= 
\mp \mathrm{sign}(p^{1})\, ,
\hspace{1cm}
\mathrm{sign}(\Im\mathfrak{m}\, Z^{2}_{\infty})  
= 
\pm \mathrm{sign}(p^{1})\mathrm{sign}(p^{2})\, .
\end{equation}

These expressions for the integration constants and the mass are valid
both for the 2- and 1-parameter families, the latter being recovered by
setting $\Im\mathfrak{m}\, Z_{\infty}=0$ everywhere. The contribution
of the monopole charge $1/g$ to the mass disappears because it is
screened.

Observe that the positivity of the mass is not guaranteed in the $-$ branch
for arbitrary values of the charges and moduli: it has to be imposed by
hand. 

Let us now study the behavior of the solution in the near-horizon
limit $r\rightarrow 0$. For $f_{\mu, s\neq 0}$ and $f_{\lambda}$ the
metric function behaves as

\begin{equation}
e^{-2U} 
\sim 
\sqrt{-2p^{1}q_{0}\left[(p^{2})^{2} -(2/g)^{2} \right]}\, \frac{1}{r^{2}}\, ,  
\end{equation}

\noindent
which corresponds to a regular horizon in both branches. The solutions will
describe regular black holes if the charges and moduli are such that
$M>0$. Observe that in the $-$ branch it is possible to chose those such that
$M=0$ with a non-vanishing entropy.

In the $f_{\mu,s=0}$ case with $p^{2}\neq 0$ the solution is only well
defined in the $+$ branch because there is no $1/r$ contribution from
the monopole in the $r\rightarrow 0$ limit and it is impossible to
satisfy the inequality
$-\eta_{\alpha\beta}\mathcal{I}^{\alpha}\mathcal{I}^{\beta}>0$ in that
limit. In this case (the $+$ branch with $p^{2}\neq 0$) we have

\begin{equation}
e^{-2U} 
\sim 
\sqrt{-2p^{1}q_{0}(p^{2})^{2}}\, \frac{1}{r^{2}} \, ,
\end{equation}

\noindent
which corresponds to a regular horizon.  

In the $f_{\mu,s=0}$ case with $p^{2}=0$ there are two possibilities: 

\begin{enumerate}
\item We can set $p^{1}=q_{0}=0$. Then, in the $r\rightarrow 0$ limit,
  $e^{-2U}$ is the moduli-dependent constant
  $2\sqrt{-2A^{1}A_{0}(A^{2})^{2}}$. There is neither horizon nor
  singularity and the solution, which is a global monopole, belongs to
  the $+$ branch (this also guarantees that the mass is positive).
\item We can keep both $p^{1}\neq 0$ and $q_{0}\neq 0$, setting
  $A^{2}=0$ and profit from the fact that, in this limit
  $\Phi^{a}\Phi^{a}$ goes to zero as $r^{2}$. The solution is only
  well defined in the $-$ branch. The metric function takes the
  constant value

\begin{equation}
e^{-2U} 
\sim 
\sqrt{+p^{1}q_{0} \frac{\mu^{4}}{g^{2}}} \, ,  
\end{equation}

\noindent
We have, as far as the metric is concerned, a global monopole solution
(as long as $M>0$), but since we need two Abelian charges switched on, namely $p^{1}$ and
$q_{0}$, the scalar fields and the gauge fields are singular at $r=0$.
As before, it is possible to tune the moduli and charges so that
$M=0$.

\end{enumerate}

The near-horizon limits of the scalars are, in the $f_{\mu, s\neq 0}$
and $f_{\lambda}$ cases 

\begin{equation}
  \begin{array}{rcl}
\Im\mathfrak{m}\tau_{\rm h}
& = &
{\displaystyle
\frac{\sqrt{-2p^{1}q_{0}\left[(p^{2})^{2} 
-(2/g)^{2} \right]}}{2\left[(p^{2})^{2} -(2/g)^{2} \right]}\, ,  
}
\\
& & \\
\Im\mathfrak{m}\, Z^{2}_{\rm h}
& = &
{\displaystyle
\frac{p^{2}}{p^{1}}\Im\mathfrak{m}\tau_{\rm h}\, ,  
}
\\
& & \\
\Im\mathfrak{m}\, Z^{a}_{\rm h}
& = &
{\displaystyle
\frac{2\Im\mathfrak{m}\tau_{\rm h}}{gp^{1}} \delta^{a}{}_{m}\frac{x^{m}}{r}\, ,  
}
\\
\end{array}
\end{equation}

\noindent
and, in the $f_{\mu, s=0}$ case with $p^{2}\neq 0$, we get the same
results up to the contribution of the monopole which disappears
(formally, $1/g=0$).


\subsection{Embedding in pure $\mathrm{SU}(2)$ EYM}

The scalars can only be trivialized for the Wu-Yang monopole
$s=\infty$.  In that case, it is easy to construct a double-extremal
black hole with constant scalars and the metric is, as usual,
Reissner-Nordstr\"om's.


\section{Multi-center SBHSs}
\label{sec-multibh}

To construct multi-center SBHSs we can use the same recipe as in the
single-center case but we need multi-center solutions of the
Bogomol'nyi equations. We start by discussing these.


\subsection{Multi-center solutions of the $\mathrm{SU}(2)$ Bogomol'nyi equations on $\mathbb{R}^{3}$}
\label{sec-multiprotogenov}

In the Abelian case, the multicenter solutions of the Bogomol'nyi equations
are associated to harmonic functions with isolated point-like singularities.
They are the seed solutions of the multi-black-hole solutions of the
Einstein-Maxwell theory
\cite{Majumdar:1947eu,kn:Pa,Perjes:1971gv,Israel:1972vx,Hartle:1972ya,Chrusciel:2005ve}
and $\mathcal{N}=2$, $d=4$ supergravities
\cite{Behrndt:1997ny,Denef:2000nb,Bates:2003vx,Bellorin:2006xr}. In the
non-Abelian case, the hedgehog ansatz is clearly inappropriate and more
sophisticated methods need to be used. Only a few explicit solutions are
known, even though solutions describing several BPS objects in equilibrium
are, on general grounds, expected to exist. For instance, there is no explicit
solution describing two BPS 't Hooft-Polyakov monopoles in equilibrium
(see however Ref.~\cite{Panagopoulos:1983yx}).

Perhaps not surprisingly, the only general families of explicit solutions
available involve an arbitrary number of Wu-Yang or Dirac monopoles embedded
in $\mathrm{SU}(2)$. The simplest of these only involve Wu-Yang monopoles and
formally, it can be obtained from solutions describing Dirac monopoles
embedded in $\mathrm{SU}(2)$ via singular gauge transformations
\cite{Popov:2004rt}, generalizing the constructions reviewed in
Appendices~\ref{app-WY} (minimal charge) and \ref{app-higherchargeWY} (higher
charge). As we have explained at length in the preceding sections, the metric
is completely oblivious to these gauge transformations and takes the same form
as in the Abelian cases. We will not study such solutions in this section.

In Refs.~\cite{Cherkis:2007jm}, using the Nahm equations
\cite{Nahm:1982jb}, Cherkis and Durcan found new solutions describing one or
two, charge 1, Wu-Yang monopoles embedded in $\mathrm{SU}(2)$ in the background of a
single BPS 't~Hooft-Polyakov monopole.\footnote{
  In Ref.~\cite{Blair:2010kz} Blair and Cherkis generated a solution describing an arbitrary number
  of charge 1 Wu-Yang monopoles in the presence of an 't Hooft-Polyakov monopole; one can easily
  generalize this solution to one describing an arbitrary number of charge $n(>0)$ Wu-Yang monopoles
  in the background of an 't Hooft-Polyakov monopole, by coalescing $n$ charge $1$ Wu-Yang monopoles.
  Needless to say, the Protogenov trick works as expected.
  For the sake of simplicity of exposition, we will not consider this more general solution in this article.
} 
We are going to use the first of them
to construct multi-center solutions of the $\overline{\mathbb{CP}}^{3}$ and
$ST[2,4]$ models of $\mathcal{N}=2$, $d=4$ SEYM. Let us review the Cherkis-Durcan
solution first: take the BPS 't~Hooft-Polyakov monopole to be located at $x^{n}=x^{n}_{0}$ and the Wu-Yang
monopole at $x^{m}=x^{m}_{1}$.  We define the coordinates relative to each of
those centers and the relative position by

\begin{equation}
r^{m}\equiv x^{m} -x^{m}_{0}\, ,
\hspace{1cm}
u^{m}\equiv x^{m}-x^{m}_{1}\, , 
\hspace{1cm}
d^{m}\equiv u^{m}-r^{m} =  x^{m}_{0} -x^{m}_{1}\, ,
\end{equation}

\noindent
and their norms by respectively, $r$, $u$ and $d$.  The Higgs field and gauge
potential of this solution (adapted to our conventions) are given by \cite{Cherkis:2007jm}

\begin{eqnarray}
\pm \Phi^{a} 
& = & 
\frac{1}{g}\delta^{a}{}_{m}
\left\{
\left[
\frac{1}{r}-\left(\mu+\frac{1}{u}\right)\frac{K}{L}
\right]
\frac{r^{m}}{r}
+\frac{2r}{uL}
\left(\delta^{mn}-\frac{r^{m}r^{n}}{r^{2}}\right) d^{n}
\right\} \, ,
\\
& & \nonumber \\
 \label{eq:48}
A^{a}
& = & 
-\frac{1}{g}\left[ 
\frac{1}{r}-\frac{\mu\mathrm{D}+2d+2u}{\mathrm{L}}
\right]\ 
\frac{\varepsilon^{a}{}_{mn}r^{m} dx^{n}}{r}
\ +\ 2\frac{\mathrm{K}}{\mathrm{L}} 
\frac{\varepsilon_{npq}d^{n}u^{p} dx^{q}}{u\mathrm{D}}\ 
\delta^{a}{}_{m}\frac{r^{m}}{r}
\nonumber \\
& & \nonumber \\
& &  
\ -\ \frac{2r}{u\mathrm{L}}\ \delta^{a}{}_{m}
\left(\delta^{mn}-\frac{r^{m}r^{n}}{r^{2}}\right)
\varepsilon_{npq} u^{p}dx^{q}\, ,
\end{eqnarray}

\noindent
where the functions $K,L,\mathrm{D}$ of $u$ and $r$ are defined by 

\begin{eqnarray}
K
& \equiv &
\left[(u+d)^{2}+r^{2}\right]\cosh\, \mu r+2r(u+d)\sinh\, \mu r\, ,\\
& & \nonumber \\
L
& \equiv & 
\left[(u+d)^{2}+r^{2}\right]\sinh\, \mu r+2r(u+d)\cosh\, \mu r\, ,\\
& & \nonumber \\
\label{eq:49}
\mathrm{D} 
& = & 
2\left( ud +u^{m}d^{m}\right)
= 
(d+u)^{2} - r^{2}\, .
\end{eqnarray}

The function $\mathrm{D}$ is clearly zero along the direction\footnote{
  This is the half of the line that joins $r=0$ to
  $u=0$ that stretches from the Dirac monopole $u=0$ to infinity in the
  direction opposite to the 't~Hooft-Polyakov monopole at $r=0$
}
$u^{m}/u=-d^{m}/d$ signaling the
possible presence of a Dirac string in Eq.~(\ref{eq:48}); that this is however
not the case is demonstrated in Ref.~\cite{Blair:2010kz}.

In the models that we are going to study, the Higgs field enters the metric in
the combination $\Phi^{a}\Phi^{a}$, which takes the value

\begin{equation}
\label{eq:fafa}
\Phi^{a}\Phi^{a}
=
\frac{1}{g^{2}}
\left\{
\left[
\frac{1}{r}-\left(\mu+\frac{1}{u}\right)\frac{K}{L}
\right]^{2}
+\frac{4|\vec{r}\times \vec{d}|^{2}}{u^{2}L^{2}}
\right\}\, .  
\end{equation}

To better understand this solution one will consider several limits:

\begin{enumerate}
\item The limit in which we take the BPS 't~Hooft-Polyakov anti-monopole
  infinitely far away, keeping the Dirac monopole at $x^{m}_{1}$: in this limit
  $d\rightarrow \infty$, $r^{m}\sim -d^{m}$ while $u$ remains finite. The
  Higgs and gauge fields take the form

\begin{eqnarray}
\label{eq:x0infty-1}
\pm \Phi^{a} 
& \sim & 
-\frac{1}{g}\delta^{a}{}_{m}
\left(\mu+\frac{1}{u}\right)
\frac{d^{m}}{d}
\, ,
\\
& & \nonumber \\
A^{a}
& \sim & 
-\frac{1}{g}
\left(1+\frac{d^{m}}{d}\frac{u^{m}}{u}\right)^{-1} 
\varepsilon_{mnp}\frac{d^{m}}{d}\frac{u^{m}}{u}d\frac{u^{p}}{u}\, .
\end{eqnarray}

The gauge field should be compared with the embedding of a Dirac monopole with
a string in the direction $-d^{m}$ into the direction
$\delta^{a}{}_{m}d^{m}T^{a}$ of the gauge group, Eqs.~(\ref{eq:Bs}) and
(\ref{eq:Diracembedded}) with $s^{m}=-d^{m}$.

\item The limit in which we take the Dirac monopole infinitely away, keeping
  the BPS 't~Hooft-Polyakov anti-monopole at $x^{m}_{0}$: In this limit
  $d\rightarrow \infty$, $u^{m}\sim d^{m}$ while $r$ remains finite. The Higgs
  and gauge fields become those of a single BPS 't~Hooft-Polyakov
  anti-monopole at $x^{m}_{0}$.

\item In the limit in which we are infinitely far away from both monopoles
  ($r\rightarrow \infty$, $u\rightarrow \infty$), which remain at a finite
  relative distance, the Higgs and gauge fields take the form

\begin{eqnarray}
\label{eq:asymp-1}
\pm \Phi^{a}
& = &   
-\left[\frac{\mu}{g}
+\mathcal{O}(|x|^{-2})\right]\delta^{a}{}_{m}\frac{x^{m}}{|x|}\, ,
\\
& & \nonumber \\
\label{eq:asymp-2}
A^{a}
& = & 
-\frac{1}{g}\varepsilon^{a}{}_{mn}\frac{x^{m}dx^{n}}{|x|^{2}}
+\frac{1}{2g}\delta^{a}{}_{m}\frac{x^{m}}{|x|}
\left(
\frac{\varepsilon_{npq}d^{n}x^{p}dx^{q}}{|x|^{2}} 
\right)\, .
\end{eqnarray}

The first term in the gauge potential is identical to that of a
Wu-Yang anti-monopole (compare with Eq.~(\ref{eq:Ameron})). This is
also the asymptotic behavior of the BPS 't~Hooft-Polyakov monopole. The
Higgs field is asymptotically covariantly constant and, in particular

\begin{equation}
\label{eq:asymp-3}
\Phi^{a}\Phi^{a}
\sim
\frac{\mu^{2}}{g^{2}}+\mathcal{O}(\frac{1}{|x|^{2}})\, .  
\end{equation}

\item The limit in which we approach the center of the BPS 't~Hooft-Polyakov
  anti-monopole $r^{m}\rightarrow 0$, $u^{m}\rightarrow d^{m}$

\begin{equation}
\Phi^{a}\Phi^{a}
\sim
\frac{1}{4g^{2}d^{2}(1+\mu d)^{2}}+\mathcal{O}(r)\, .  
\end{equation}

This limit is finite and only vanishes when the Dirac monopole is taken to
infinity $d\rightarrow \infty$.  

For finite values of $d$, Eq.~(\ref{eq:fafa}) says that $\Phi^{a}\Phi^{a}$ can
only vanish along the line that stretches from $r=0$ to $u=0$ so
$\vec{r}\times \vec{d}=0$.  Substituting $r^{m}=\alpha d^{m}$ in
$\Phi^{a}\Phi^{a}$ we get a function of $\alpha$ and of the parameter $\mu d$.
%
\begin{figure}[t]
  \centering
  \begin{tikzpicture}[scale=0.9]
    \begin{axis}[ 
        xlabel=$\mu d$,
        ylabel=\rotatebox{-90}{$\textstyle{r(d)\over d}$},      
        xmin=0,
        xmax=2.1,
        ymin=0,
        ymax=1.1,
        axis x line*=bottom,     
        axis y line*=left,            
        xtick align=center,
        ytick align=center
      ]
      \addplot [color=black] table[x=d,y=x1] {BnC-ZerosHiggs.dat};
    \end{axis}
   \end{tikzpicture}
  \caption{\label{fig:BnCZerosHiggs}
                The zeros of the Higgs density as measured by $r$ as a function of the dimensionless separation $\mu d$.
               }
\end{figure}
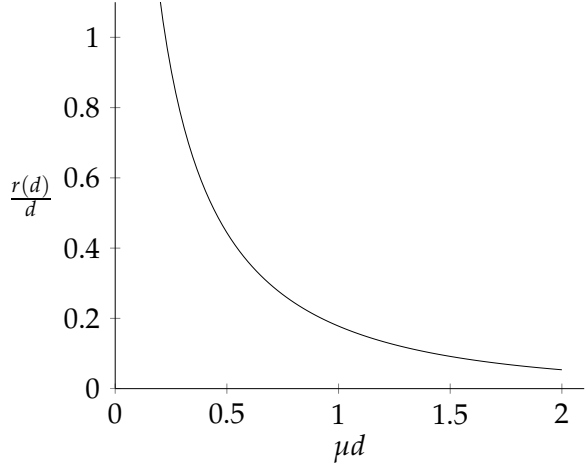
%
Plotting the functions of $\alpha$ for different values of $\mu d$ we find
that they have a single zero, which is also a local minimum. At this minimum the
second derivative does not vanish, and therefore, there,
$\Phi^{a}\Phi^{a}\sim\mathcal{O}(r^{2})$, as in the single-monopole case.
However, the value of this second derivative depends on the direction.

\item The limit in which we approach the singularity of the Dirac monopole
  $u^{m}\rightarrow 0$, $r^{m}\rightarrow -d^{m}$

\begin{equation}
\label{eq:nearDirac}
\Phi^{a}\Phi^{a}
\rightarrow
\frac{1}{g^{2}}
\left\{
\frac{1}{u^{2}}
+\left(\frac{1}{d}-\mu \right)\frac{1}{u}
\right\} +\mathcal{O}(1)\, .  
\end{equation}

\end{enumerate}


\subsubsection{Growing Protogenov hair}

As we have argued in Sec.~(\ref{sec:PTrick}) we can add a Protogenov hair parameter $s$ to the
Cherkis \& Durcan solution by simply replacing the argument $\mu r$ of
the hyperbolic sines and cosines in the functions $K$ and $L$ by the
shifted on $\mu r+s$. We do not need to write explicitly the solution,
but we do need to reconsider the different limits studied for the
$s=0$ case:

\begin{enumerate}
\item In the limit in which we take the BPS 't~Hooft-Polyakov-Protogenov
  anti-monopole infinitely away, keeping the Dirac monopole at $x^{m}_{1}$ the
  Higgs and gauge fields become, to leading order, those of the Dirac monopole
  with the Dirac string in the direction $-d^{m}$, as in the $s=0$ case
  (See Eqs.~(\ref{eq:x0infty-1}) and (\ref{eq:48})).

\item In the limit in which we take the Dirac monopole infinitely away,
  keeping the BPS 't~Hooft-Polyakov-Protogenov anti-monopole at $x^{m}_{0}$
  the Higgs and gauge fields become those of a single BPS
  't~Hooft-Polyakov-Protogenov anti-monopole at $x^{m}=x^{m}_{0}$ (the first
  two equations (\ref{eq:Protogenovsolutions})).

\item In the limit in which we are infinitely far away from both monopoles
  ($r\rightarrow \infty$, $u\rightarrow \infty$), which remain at a finite
  relative distance, the Higgs and gauge fields take the same form as in the
  $s=0$ case, Eqs.~(\ref{eq:asymp-1}-\ref{eq:asymp-3}). 

\item The limit in which we approach the singularity of the BPS
  't~Hooft-Polyakov-Protogenov anti-monopole $r^{m}\rightarrow 0$,
  $u^{m}\rightarrow d^{m}$ (for $s\neq 0$)

\begin{eqnarray}
\pm \Phi^{a}
& \sim &
\frac{1}{g}\delta^{a}{}_{m}
\left[ 
\frac{1}{r} - \left(\mu+\frac{1}{d}\right)\coth{s} +\mathcal{O}(r)
\right]
\frac{r^{m}}{r}\, ,
\\
& & \nonumber \\
\Rightarrow \Phi^{a}\Phi^{a}
& \sim &
\frac{1}{g^{2}r^{2}}
+\mathcal{O}\left(\frac{1}{r}\right)\, ,
\end{eqnarray}

\noindent
which is similar to the behaviour near the Dirac monopole as in
Eq.~(\ref{eq:nearDirac}) (with $u$ replaced by $r$).

\item The limit in which we approach the singularity of the Dirac monopole
  $u^{m}\rightarrow 0$, $r^{m}\rightarrow -d^{m}$ we have the same behavior as
  in the $s=0$ case Eq.~(\ref{eq:nearDirac}).

\end{enumerate}

The solutions with Protogenov hair have another limit, namely the one in which $s\rightarrow\infty$;
this case will be studied separately.


\subsubsection{The $s\rightarrow \infty$ limit solution}

In this limit we get a solution that describes the same Dirac monopole
together with a ($\mu\neq 0$) Wu-Yang anti-monopole:\footnote{
  One can see fairly easily that in the limiting solution one can, as far as the Bogomol'nyi
  equations are concerned, allow for $\mu$ to be negative; for finite values of $s$ this
  is impossible. 
}

\begin{eqnarray}
\pm \Phi^{a}
& = &
\frac{1}{g}\delta^{a}{}_{m}
\left[-\mu +\frac{1}{r} -\frac{1}{u}
\right]
\frac{r^{m}}{r}\, ,
\\
& & \nonumber \\
A^{a}
& = &
\frac{1}{g} \frac{\varepsilon^{a}{}_{mn}r^{m}dx^{n}}{r^{2}}
+
\frac{1}{g} \frac{\varepsilon_{npq}d^{n}u^{p}du^{q}}{u(ud+u^{r}d^{r})} 
\delta^{a}{}_{m}\frac{r^{m}}{r}\, .
\end{eqnarray}

This solution is a particular example of a more general family
describing an arbitrary number of Dirac monopoles in the background of
a Wu-Yang anti-monopole. These solutions can be obtained from a
solution describing only Dirac monopoles embedded in $\mathrm{SU}(2)$
via a singular gauge transformation that only removes the Dirac string
of one of them, which becomes the Wu-Yang anti-monopole. The general
family of solutions can be written in the form:

\begin{equation}
\Phi = \Phi_{\rm WY} +H U\, ,
\hspace{1cm}
A = A_{\rm WY} +C U\, ,    
\end{equation}

\noindent
where $U$ is the $\mathrm{SU}(2)$ (and $\mathfrak{su}(2)$) matrix
defined in Eq.~(\ref{eq:U}) and where $\Phi_{\rm WY}$ and $A_{\rm WY}$
are the Higgs and Yang-Mills fields of a Wu-Yang monopole, given,
respectively, by

\begin{equation}
\mp \Phi_{\rm WY}
 = 
\frac{1}{2g}
\left[-\mu +\frac{1}{r}\right] U\, ,
\end{equation}

\noindent
and by Eq.~(\ref{eq:Ameron}) and where $H$ is a function and $C$ a
1-form on $\mathbb{R}^{3}$. If we substitute into the Bogomol'nyi
equations (\ref{eq:Beqs}) and use, on the one hand, that they are
satisfied by the pair $A_{\rm WY},\Phi_{\rm WY}$, and, on the other
hand, that $U$ is covariantly constant with the connection $A_{\rm
  WY}$ we arrive at the Dirac monopole equation

\begin{equation}
\label{eq:dCdH}
dC = \star_{(3)}dH\, .  
\end{equation}

\noindent
The integrability condition of this equation is $d\star_{(3)}dH=0$ so
$H$ is any harmonic function. We can choose it to have isolated poles at
the points $x^{m}=x^{m}_{i}$ $i=1,\cdots, N$

\begin{equation}
  H = \sum_{i} \frac{p_{i}}{2 u_{i}}\, ,
  \hspace{1cm}
  u_{i}^{m} \equiv x^{m}-x^{m}_{i}\, ,
\end{equation}

\noindent
in which case $C$ is the 1-form potential of $N$ Dirac monopoles with charges $p_{i}$
which can be constructed by summing over the potentials of each individual
monopole:

\begin{equation}
C= \sum C_{i}\, ,
\hspace{1cm}
dC_{i} = \star_{(3)} d \frac{p_{i}}{2 u_{i}}\, .
\end{equation}

The expression for each of the $C_{i}$ is of the form Eq.~(\ref{eq:Bs}) where
we can, in principle, choose the direction $s^{m}_{i}$ of each Dirac string
independently:

\begin{equation}
C_{i} = 
\frac{p_{i}}{2}\left(1-\frac{s_{i}^{m}}{s_{i}}\frac{u_{i}^{m}}{u_{i}}\right)^{-1} 
\varepsilon_{mnp}\frac{s_{i}^{m}}{s_{i}}\frac{u^{m}_{i}}{u_{i}}d\frac{u_{i}^{p}}{u_{i}}\, ,
\hspace{.5cm}
\mbox{(no sum over $i$)}.   
\end{equation}

This solution of the Yang-Mills-Higgs system shares two important
properties with the original Wu-Yang monopole and which are related to
the fact that they are related to Abelian embeddings by singular gauge
transformations:

\begin{enumerate}
\item Both $\Phi$ and $D\Phi$ are proportional to $U$:

  \begin{equation}
\Phi = \left(-\frac{\mu}{2g}+\frac{1}{2gr} +H\right) U\, ,
\hspace{1cm}
D\Phi = d\left(-\frac{\mu}{2g}+\frac{1}{2gr} +H\right) U\, ,
  \end{equation}

\noindent
and, therefore, commute with each other, so the Higgs current vanishes and the
gauge field is, by itself, a solution of the pure Yang-Mills theory.

\item The gauge field strength is also proportional to $U$, the coefficient
  being the field strength of an Abelian gauge field:

  \begin{equation}
  F(A) = d(B+C) U\, ,  
  \end{equation}

\noindent
which implies that the energy-momentum tensors are related as in the
single-center case.

\end{enumerate}

These solutions can be generalized even further, by allowing the the
charge of the ``original'' Wu-Yang monopole at $r=0$ to be $n/g$ (that
is: using the generalization of the Wu-Yang monopole due to Bais
\cite{Bais:1976fr} which is studied in
Appendix~\ref{app-higherchargeWY}).  If we now substitute into the
Bogomol'nyi equations (\ref{eq:Beqs}) the ansatz

\begin{equation}
\Phi = \Phi_{(n)} +H U_{(n)}\, ,
\hspace{1cm}
A =A_{(n)} +C U_{(n)}\, ,    
\end{equation}

\noindent
where $U_{(n)},A_{(n)}$ and $\Phi_{(n)}$ are given, respectively, in
Eqs.~(\ref{eq:Un}),(\ref{eq:Ameronn}) and (\ref{eq:Phimeronn}), $H$ is
a function and $C$ a 1-form on $\mathbb{R}^{3}$, and use that they are
satisfied by the pair $A_{(n)},\Phi_{(n)}$ and that $U_{(n)}$ is
covariantly constant with the connection $A_{(n)}$, we arrive again at
the Dirac monopole equation (\ref{eq:dCdH}).

Since all these solutions are related to Abelian embeddings, they
contribute to the black-hole solutions as the Abelian solutions. We
will not consider them in what follows.


\subsection{Embedding in the  $\mathrm{SU}(2)$-gauged 
$\overline{\mathbb{CP}}^{3}$ model}

We can use the Cherkis \& Durcan solution of the $\mathrm{SU}(2)$ Bogomol'nyi
equations reviewed in the previous section as a seed solution for a
multicenter solution of $\mathcal{N}=2$, $d=4$ SEYM, adding the same harmonic
functions as in the single-center case ($\mathcal{I}^{0},\mathcal{I}_{0}$) or
a generalization with poles at the locations of the monopoles
$r=0$\footnote{The location of the BPS 't~Hooft-Polyakov anti-monopole is not
  completely clear: it is sometimes argued that the center of the monopole is
  the point at which the Higgs vanishes and the full gauge symmetry is
  restored. As we have discussed, that point is not $r=0$. We could try to
  place the poles of the harmonic functions at that point, but, given that its
  location is not known analytically and the expansion of $\Phi^{a}\Phi^{a}$
  around it is difficult to compute, we will not try to do that here.} and
$u=0$. More explicitly, we take

\begin{equation}
\begin{array}{rcl}
\mathcal{I}^{0} 
& = & 
A^{0}+{\displaystyle\frac{p^{0}_{r}/\sqrt{2}}{r}}
+{\displaystyle\frac{p^{0}_{u}/\sqrt{2}}{u}}\, ,
\\
& & \\  
\mathcal{I}_{0} 
& = & 
A_{0}+{\displaystyle\frac{q_{r,0}/\sqrt{2}}{r}}
+{\displaystyle\frac{q_{u,0}/\sqrt{2}}{u}}\, ,
\\
& & \\
\mathcal{I}^{i}
& = & 
\mp\sqrt{2} \Phi^{i}(r,u)\, ,
\\
& & \\
\mathcal{I}_{i}
& = & 
0\, ,
\end{array}
\end{equation}

\noindent
where $\Phi^{i}(r,u)$ is the Higgs field of the Cherkis \& Durcan
solution. The metric and scalar fields take the form

\begin{eqnarray}
e^{-2U}
& = &
\tfrac{1}{2}(\mathcal{I}^{0})^{2} +2(\mathcal{I}_{0})^{2} -\Phi^{i}\Phi^{i}\, ,
\\
&  & \nonumber \\
Z^{i}
& = & 
\frac{\mp\sqrt{2}\Phi^{i}}{\mathcal{I}^{0}+2i\mathcal{I}_{0}}\, .  
\end{eqnarray}

The normalization of the metric and scalars at infinity leads to the
same relations between the integration constants $A^{0},A_{0},\mu$ and
the physical constants $Z_{\infty},g$ as in the single-center case, namely

\begin{equation}
\tfrac{1}{\sqrt{2}}A^{0}+\sqrt{2}i A_{0}
=
\frac{Z^{*}_{\infty}}{|Z_{\infty}|}\frac{1}{\sqrt{1 -|Z_{\infty}|^{2}}}\, ,
\hspace{1cm}
\mu
=
\frac{|Z_{\infty}|}{\sqrt{1 -|Z_{\infty}|^{2}}} g\, .
\end{equation}

The integrability conditions of Eq.~(\ref{eq:omega}) are, in this case,

\begin{equation}
\mathcal{I}_{0}\partial_{\underline{m}}\partial_{\underline{m}}\mathcal{I}^{0} 
-
\mathcal{I}^{0} \partial_{\underline{m}}\partial_{\underline{m}}\mathcal{I}_{0} =0\, ,
\end{equation}

\noindent
and lead to the following relations between the integration constants:

\begin{eqnarray}
\label{eq:omega=0nonA2centers}
A^{0}(q_{r,0}+q_{u,0})-A_{0}(p^{0}_{r}+p^{0}_{u}) 
& = &
0\, ,
\\
& & \nonumber \\ 
J-\tfrac{1}{\sqrt{2}}d(A^{0}q_{u,0}-A_{0}p^{0}_{u})
& = &
0\, ,
\end{eqnarray}

\noindent
where we have defined the constant

\begin{equation}
\label{eq:J}
J\equiv p_{r}^{0}q_{u,0}-q_{r,0}p^{0}_{u}\, .  
\end{equation}

The first equation is equivalent to Eq.~(\ref{eq:omega=0nonA}) for the
total charges and the second equation determines the relative distance
$d$ in terms of $J$ and $A^{0}q_{u,0}-A_{0}p^{0}_{u}$ provided that
$J\neq 0$. When that is the case, the solution is not static and has
an angular momentum $J$ directed along the line that joins the
monopoles $J^{m}=Jd^{m}/d$. The corresponding 1-form $\omega$ can be
constructed by the standard procedure of the Abelian case. However,
since this complicates the analysis of the regularity of the
solutions, we will stick to the static case and require $J=0$. 

In order to have regular solutions, the charges at each center must be
chosen as in the corresponding single-center case: since there is an
Abelian monopole at $u=0$, we must switch on either $p^{0}_{u}$ or
$q_{u,0}$ to have a regular horizon there. We can treat them both as
non-vanishing with no loss of generality. Then, there are two
possibilities:

\begin{description}
\item[I. $p_{r}^{0}=q_{r,0}=0$:] Only for $s=0$ ('t~Hooft-Polyakov
  anti-monopole at $r=0$) has the solution a chance of being regular
  at $r=0$. Solving Eq.~(\ref{eq:omega=0nonA2centers}) the solution
  can be written in the form

\begin{eqnarray}
\label{eq:CP32centermetric}
e^{-2U}
& = &
\frac{1}{1-|Z_{\infty}|^{2}}H^{2} -\Phi^{i}\Phi^{i}\, ,
\\
&  & \nonumber \\
\label{eq:CP32centerscalars}
Z^{i}
& = & 
\frac{2\beta}{p^{0}+2iq_{0}} \frac{\Phi^{i}}{H}\, ,  
\end{eqnarray}

\noindent
where $H$ is the harmonic function 

\begin{equation}
H\equiv 1+\frac{\beta}{u}\, ,
\hspace{.7cm}
\beta^{2}
=
(1-|Z_{\infty}|^{2})W_{\rm RN}(\mathcal{Q}_{u})/2\, ,
\hspace{.7cm}
 W_{\rm RN}(\mathcal{Q}_{u})
\equiv 
\tfrac{1}{2}(p^{0}_{u})^{2} +2(q_{u,0})^{2}\, .
\end{equation}

The free parameters of this solution are the charges
$p^{0}_{u}$, $q_{u,0}$ and the single modulus $|Z_{\infty}|$.

Studying the $u\rightarrow 0$ limit we find a black hole with entropy

\begin{equation}
S_{u}/\pi 
=
\tfrac{1}{2}
W_{\rm RN}(\mathcal{Q}_{u})
-\frac{1}{g^{2}}\, ,  
\end{equation}

\noindent
as in the corresponding single-center case. 

In the $r\rightarrow 0$ limit $e^{-2U}$ is constant. The positivity of
the constant is guaranteed if $S_{u}$ is positive. 
The total entropy of the solution is just the entropy of
the black hole at $u=0$ and the Dirac monopole does contribute to it.

The mass of the solution, expressed in terms of the independent
parameters of the solution, $p^{0}_{u}$, $q_{u,0}$ and $|Z_{\infty}|$ takes the form

\begin{eqnarray}
M
& = & 
M_{r}+M_{u}\, ,
\\
& & \nonumber \\
M_{r}
& = & 
-M_{\rm monopole}\, ,
\\
M_{u}
& = & 
\sqrt{\tfrac{1}{2}\frac{W_{RN}(\mathcal{Q}_{u})}{1-|Z_{\infty}|^{2}}}+M_{\rm monopole}\, ,
\end{eqnarray}

\noindent
where $M_{\rm monopole}$ is given by Eq.~(\ref{eq:massmonopole}). The
contributions of the monopole and the 't Hooft-Polyakov monopole to the mass cancel
each other. 

\item[II. $p_{r}^{0}$ or $q_{r,0}\neq 0$] We can treat both charges as
  non-vanishing with no loss of generality. Solving
  Eqs.~(\ref{eq:omega=0nonA2centers}) and (\ref{eq:J}), we can write
  the solution as in Eqs.~(\ref{eq:CP32centermetric}) and
  (\ref{eq:CP32centerscalars}) where, now,

\begin{equation}
  \begin{array}{rcl}
H 
& \equiv &
{\displaystyle
1+\frac{\beta_{r}}{r}+\frac{\beta_{u}}{u}\, ,
}
\hspace{1cm}
\beta_{r,u}^{2}
=
(1-|Z_{\infty}|^{2})W_{\rm RN}(\mathcal{Q}_{r,u})/2\, ,
\\
& & \\
W_{\rm RN}(\mathcal{Q}_{r,u})
& \equiv &
\tfrac{1}{2}(p^{0}_{r,u})^{2} +2(q_{r,u,0})^{2}\, .
\end{array}
\end{equation}

The free parameters of this solution are the charges
$p^{0}_{u}$, $q_{u,0}$ and $|Z_{\infty}|$ and either $p^{0}_{r}$ or $q_{r,0}$, since
they must be proportional to those of the other center. The areas of
each of the horizons are as in the single-center case. In particular,
the BPS 't~Hooft-Polyakov monopole ($s=0$) does not contribute to the
entropy of the $r=0$ center. The mass is given by 

\begin{eqnarray}
M
& = & 
M_{r}+M_{u}\, ,
\\
& & \nonumber \\
M_{r}
& = & 
\sqrt{\tfrac{1}{2}\frac{W_{RN}(\mathcal{Q}_{r})}{1-|Z_{\infty}|^{2}}}-M_{\rm monopole}\, ,
\\
M_{u}
& = & 
\sqrt{\tfrac{1}{2}\frac{W_{RN}(\mathcal{Q}_{u})}{1-|Z_{\infty}|^{2}}}+M_{\rm monopole}\, ,
\end{eqnarray}

\noindent
and the contributions of the monopole and anti-monopole cancel each
other.  In the $s\rightarrow \infty$ limit it can be easily seen that
the solution is completely regular everywhere ($e^{-2U}$ only vanishes
at $r=0$ and $u=0$) if the Abelian charges as chosen so that the
horizons are regular. This guarantees that all the terms in $e^{-2U}$
are positive. For finite $s$ this is more difficult to proof
analytically, but, since the Higgs field has a better behavior than in
the $s\rightarrow \infty$ case, it is reasonable to expect that it
will also be true. We have checked numerically that this is so in
several examples.

\end{description}


\subsection{Embedding in the $\mathrm{SU}(2)$-gauged 
$\mathrm{ST}[2,4]$ model}


The metric and scalar fields of the solution are now given by

\begin{eqnarray}
e^{-2U}
& = &
2\sqrt{-2\mathcal{I}^1\mathcal{I}_{0}
[(\mathcal{I}^{2})^{2}-2\Phi^{a}\Phi^{a}]
}\, ,
\\
& & \nonumber \\
Z^{1}
& \equiv &
\tau
= 
i\frac{e^{-2U}}{2[(\mathcal{I}^{2})^{2}-2\Phi^{a}\Phi^{a}]}\, ,
\hspace{1cm}
Z^{2} 
= 
\frac{\mathcal{I}^{2}}{\mathcal{I}^{1}} \tau\, ,
\hspace{1cm}
Z^{a} 
 = 
\frac{\sqrt{2}\Phi^{a}}{\mathcal{I}^{1}} \tau\, ,
\end{eqnarray}

\noindent
where $\Phi^{a}$ is the Higgs field of the Cherkis \& Durcan solution
(deformed with the Protogenov hair parameter $s$) and where the harmonic
functions $\mathcal{I}^{1}$, $\mathcal{I}^{2}$ and $\mathcal{I}_{0}$ are allowed to
have poles at $r=0$ and $u=0$:

\begin{equation}
\begin{array}{rclrcl}
\mathcal{I}^{1} 
& = & 
A^{1}+{\displaystyle\frac{p^{1}_{r}/\sqrt{2}}{r}}
+{\displaystyle\frac{p^{1}_{u}/\sqrt{2}}{u}}\, , 
\hspace{.6cm}
&
\mathcal{I}^{2} 
& = & 
A^{2}+{\displaystyle\frac{p^{2}_{r}/\sqrt{2}}{r}}
+{\displaystyle\frac{p^{2}_{u}/\sqrt{2}}{u}}\,  ,
\\
& & & & & \\
\mathcal{I}_{0}
& = & 
A_{0}+{\displaystyle\frac{q_{r,0}/\sqrt{2}}{r}}
+{\displaystyle\frac{q_{u,0}/\sqrt{2}}{u}}\, .
& & & 
\\
\end{array}
\end{equation}

As in the $\overline{\mathbb{CP}}^{3}$ case, the Abelian charges at
each center must be chosen with the same criteria as in the
corresponding single-center case. This means, in particular, that the
Abelian charges at $u=0$, $p^{1}_{u},q_{u,0}$ must be
non-vanishing. $p^{2}_{u}$ may need to be activated, depending on the
branch we are considering. At $r=0$, for $s\neq 0$ we get exactly the
same possibilities, but, for $s=0$ there are two possibilities:

\begin{enumerate}
\item $p^{1}_{r}$, $q_{r,0},p^{2}_{r}$ non-vanishing. We find a black
  hole at $r=0$ in the $+$ branch. 
\item $p^{1}_{r}=q_{r,0}=p^{2}_{r}=0$. $e^{-2U}$ is a complicated
  $d$-dependent constant in the $r=0$ limit and we get a global monopole.
\end{enumerate}

Here we find an important difference with the single-center case, due
to the fact that $\Phi^{a}\Phi^{a}$ is a finite constant in the
$r\rightarrow 0$ limit instead of going to zero as $r^{2}$: there is
no solution with $p^{1}_{r}q_{r,0}\neq 0$ and $p^{2}_{r}=0$. In order
to have such a global monopole solution with $p^{1}q_{0}\neq 0$ and
$p^{2}=0$ in equilibrium with the monopole at $u=0$ one may try to
place those charges at the point at which $\Phi^{a}\Phi^{a}=0$, but
the resulting solution may not be well defined there because the limit of the metric function
depends on the direction from which we approach that point.

The entropy of the solution is the sum of the entropies of both
centers (vanishing for global monopoles). As in the
$\overline{\mathbb{CP}}^{3}$ case, the monopole at each center does
contribute to the center entropy (except for global monopoles). The
contributions of the monopole and anti-monopole to the mass cancel each other:

\begin{equation}
M  
 = 
\tfrac{1}{4}
\frac{\chi_{\infty}}{|\Im\mathfrak{m}\tau_{\infty}|}
|p^{1}_{u}+p^{1}_{r}|
+
\frac{1}{2\chi_{\infty}}|q_{u,0}+q_{r,0}|
\pm\tfrac{1}{2} 
\frac{|\Im\mathfrak{m}\tau_{\infty}\Im\mathfrak{m}\, Z^{2}_{\infty}|}{\chi_{\infty}}
|p^{2}_{u}+p^{2}_{r}|
\, .
\end{equation}


\section{Conclusions}
\label{sec-conclusions}


In this article we have discussed the construction of supersymmetric
multi-object solutions in $\mathcal{N}=2$, $d=4$ EYM theories,
specifically in the so-called $\overline{\mathbb{CP}}^{n\geq 3}$ and
$\mathrm{ST}[2,n]$ models. These models were chosen due to their
workability, the fact that they allow for a $\mathrm{SU}(2)$ gauging
and (in the second case) for their stringy origin. Starting with a
deformation of the solutions to the $\mathrm{SU}(2)$ Bogomol'nyi
equation found by Cherkis and Durcan that adds to the 't
Hooft-Polyakov monopole \textit{Protogenov hair}, we have been able to
construct \textit{bona fide} two-center solutions. These solutions
describe a Dirac monopole embedded in $\mathrm{SU}(2)$ in the presence
of either a global monopole (the supergravity solution corresponding
to the 't Hooft-Polyakov monopole) or a non-Abelian black hole (a
supergravity solution with an 't Hooft-Polyakov-Protogenov
monopole). In order to make the comparison with the single-object case
easier, we included a detailed discussion of the embeddings of the
spherically symmetric solutions to the $\mathrm{SU}(2)$ Bogomol'nyi
equations into the two models, and expressed the whole solution in
terms of charges and moduli of the physical fields.

The constructed solutions are all static. It would be very interesting
to study dyonic solutions and to see how this interplays with the
Denef constraint; the stumbling block in this respect is not so much
the Bogomol'nyi equation as the equation (\ref{eq:DDILd2}); for the
moment the only general solution we know of is to take
$\mathcal{I}_{\Lambda}\sim\mathcal{I}^{\Lambda}$ in the gauged
directions, but this automatically solves the Denef constraint. The
only case for which we can find non-trivial dyonic solutions is for
the multi-Wu-Yang solutions, or if you like the $s\rightarrow\infty$
limit of the deformed Cherkis and Durcan's solution; we refrain from
discussing these solutions here as, due to gauge invariance, even
taking into account the singular gauge transformation, the
restriction coming from the Denef constraint is basically the one
corresponding to the Abelian theory.

A natural question that follows from the results presented here and in
Refs.~\cite{Huebscher:2007hj,Meessen:2008kb,Hubscher:2008yz} is
whether we could use a charge $k$ $\mathrm{SU}(2)$ monopole to
construct globally regular solutions; the answer is yes: observe that
the construction of globally regular solutions in
Sec.~(\ref{sec-single}) hinges exclusively but crucially on the fact
that the used monopole solution is regular and is such that
$\Phi^{a}\Phi^{a}\leq
\lim_{|\vec{x}|\rightarrow\infty}\Phi^{a}\Phi^{a}$.  A charge-$k$
monopole may be rather difficult to construct but the regularity is
guaranteed and also the last needed ingredient is known to be
satisfied: indeed, using the Bogomol'nyi equation (\ref{eq:Beqs}) one
can show that

\begin{equation}
\label{eq:3}
\partial_{\underline{m}}\partial_{\underline{m}}\ \Phi^{a}\Phi^{a}
 \; =\; 
F^{a}{}_{\underline{m}\underline{m}}F^{a}{}_{\underline{m}\underline{m}}\; \geq\; 0 \; .
\end{equation}

This equation together with the Hopf maximum principle and the
regularity, implies that the function $\Phi^{a}\Phi^{a}$ is bounded from
above by its value on the sphere at infinity, which is exactly what one
needs.

As was said in the introduction, the creation and study of non-Abelian
solutions to $d=4$ supergravity theories is in its infancy and this holds
doubly so for the higher dimensional theories. One possible reason is
that the structure of supersymmetric solutions to higher supergravities (see
{\em e.g.\/} Refs.~\cite{Cariglia:2004kk,Bellorin:2007yp}) is more
entangled than the one given in the recipe in
Section~\ref{sec-recipe}. For example, naively one would expect that
Kronheimer's link of monopoles on $\mathbb{R}^{3}$ to instantons on
GH-spaces, would carry over to the supersymmetric solutions as in
$d=4$ the base space is $\mathbb{R}^{3}$ and that in $d=5$ must be
hyper-K\"ahler; {\em i.e.\/} one would expect the instanton equation
to show up in the recipe for cooking up 5-dimensional supersymmetric
solutions. Perhaps it does, but it definitely is not obvious where and
how it is making its appearance in such a clear-cut manner as in
$d=4$.

The 4- and 5-dimensional EYMH theories are, however, related by
dimensional reduction/oxidation, whence the solutions to the cubic
models presented in this article can be oxidized to 5-dimensions and
can be studied with the hope of unraveling the structure of
5-dimensional supersymmetric solutions. Work along these lines is in
progress.


\section*{Acknowledgments}


The authors wish to thank A.~Giacomini and F.~Canfora for interesting
discussions and collaboration in the early stages of this article; PM
also wishes to thank D.~Rodr\'{\i}guez-G\'omez and J.~Schmude for
interesting discussions. PB wishes to thank CERN's Theory Division for hospitality.

This work has been supported in part by the Spanish Ministry of
Science and Education grant FPA2012-35043-C02 (-01 {\&} -02), the
Centro de Excelencia Severo Ochoa Program grant SEV-2012-0249, the
Comunidad de Madrid grant HEPHACOS S2009ESP-1473 and the Spanish
Consolider-Ingenio 2010 program CPAN CSD2007-00042 and EU-COST action
MP1210 ``The String Theory Universe''.  The work was further supported
by the JAE-predoc grant JAEPre 2011 00452 (PB), the Ram\'on y Cajal
fellowship RYC-2009-05014 (PM) and the \textit{Severo Ochoa}
pre-doctoral grant SVP-2013-067903 (PF-R).  TO wishes to thank
M.M.~Fern\'andez for her permanent support.

\appendix

\section{The $\mathrm{SU}(2)$ Lorentzian meron}
\label{app-lorentzianmeron}

A Lorentzian meron is a classical solution to the pure $\mathrm{SU}(2)$
(Lorentzian) Yang-Mills theory such that the 1-form gauge field $A$
defining it, is proportional to a pure-gauge configuration, which in our
conventions would be $\tfrac{1}{g}dU U^{-1}$ where $U(x)
\in \mathrm{SU}(2)$.  In Ref.~\cite{Canfora:2012ap} $U(x)$ was chosen to be of
the \textit{hedgehog form}

\begin{equation}
\label{eq:U}
U \equiv 2\frac{x^{m}}{r}\delta_{m}^{a} T_{a}\, ,   
\hspace{1cm}
U^{\dagger}=U^{-1}= -U\, ,
\,\,\,\,
\Rightarrow
U^{2}=-\mathbbm{1}_{2\times 2}\, .
\end{equation}

\noindent
and it was shown that $A$ solves the Yang-Mills equations if the
proportionality coefficient is $1/2$, that is

\begin{equation}
\label{eq:Ameron}
A = \frac{1}{2g}dU U^{-1} = -\frac{1}{gr^{2}} \varepsilon^{a}{}_{mn}x^{m}dx^{n}
T_{a}\, .  
\end{equation}

As we will see, this gauge field is nothing but the gauge field of the
Wu-Yang $\mathrm{SU}(2)$ monopole given in Eq.~(\ref{eq:WYmonopole}).

Since the field strength of a pure gauge configuration vanishes, we find that
$F(A)$ can be written in these two specially simple ways which we will use in
Appendix~\ref{app-skyrme}:

\begin{equation}
\label{eq:F(A)}
F(A) = \tfrac{1}{2}dA = g[A,A] = \star_{(3)} d \frac{1}{2gr}\, U\, ,
\end{equation}

Now we can write the non-Abelian field strength $F(A)$ in terms of
$F(B)$, where $F(B)$ is the field strengths of the Dirac monopole of
unit charge Eq.~(\ref{eq:F(B)}) that we will review in the next
section

\begin{equation}
F(A) = F(B) \, U\, ,
\hspace{1cm}
F(B)= \star_{(3)} d \frac{1}{2gr}\, ,
\end{equation}

\noindent
and the energy-momentum tensor of $A$ in terms of that of $B$

\begin{equation}
\label{eq:emtensorsrelated}
T_{\mu\nu}(A)
=
-\tfrac{1}{2}\mathrm{Tr}[F_{\mu\rho}(A)F_{\nu}{}^{\rho}(A)-\tfrac{1}{4}\eta_{\mu\nu}F^{2}(A)]
=
F_{\mu\rho}(B)F_{\nu}{}^{\rho}(B)-\tfrac{1}{4}\eta_{\mu\nu}F^{2}(B)  
=
T_{\mu\nu}(B)\, .
\end{equation}


\section{The Wu-Yang $\mathrm{SU}(2)$ monopole}
\label{app-WY}

The Wu-Yang $\mathrm{SU}(2)$ monopole \cite{Wu:1967vp} is a solution
of the $\mathrm{SU}(2)$ Yang-Mills theory that can be obtained from
the embedding of the Dirac monopole in $\mathrm{SU}(2)$ via a singular
gauge transformation (see, \textit{e.g.}~Ref.~\cite{Shnir:2005xx} and
references therein). To fix our conventions, it is convenient to start
by reviewing the Wu-Yang construction of the Dirac monopole
\cite{Wu:1975es}.


\subsection{The Dirac monopole}

The $\mathrm{U}(1)$ field of the Dirac monopole, that we will denote
by $B$ is defined to satisfy the Dirac monopole equation\footnote{This
  equation is just the Abelian version of the Bogomol'nyi equation.},
which can be written in several forms:

\begin{equation}
\label{eq:F(B)}
F(B)\equiv dB = \star_{(3)} d \frac{1}{2gr} = -\frac{1}{2g} d\Omega^{2}\, ,
\hspace{1cm}
2\partial_{[m}B_{n]} = -\frac{1}{2g}\varepsilon_{mnp}\frac{x^{p}}{r^{3}}\, ,
\end{equation}

\noindent
where $d\Omega^{2}$ is the volume 2-form of the round 2-sphere of unit
radius

\begin{equation}
d\Omega^{2}
= 
- \tfrac{1}{2}  \varepsilon_{mnp} \frac{x^{m}}{r} d\frac{x^{n}}{r} \wedge
d\frac{x^{p}}{r}
= 
\sin{\theta}d\theta \wedge d\varphi\, . 
\end{equation}

\noindent
The value of the magnetic charge has been set to $g^{-1}$ and it is
the minimal charge allowed if the unit of electric charge is $g$.

The above equation does not admit a global regular solution.

\begin{equation}
\label{eq:Bpm}
B^{(\pm)} = - \frac{1}{2g} (\cos{\theta} \mp 1)d\varphi\, ,  
\end{equation}

\noindent
are local solutions regular everywhere except on the negative (resp.~positive)
$z$ axis (the Dirac strings).  A globally regular solution can be constructed
by using $B^{\pm}$ in the upper (lower) hemisphere and using the gauge
transformation 

\begin{equation}
B^{(+)} -B^{(-)} = -d\left(\frac{1}{g} \varphi\right)\, ,   
\end{equation}

\noindent
to relate them in the overlap region. If the gauge group is $\mathrm{U}(1)$
where the radius of the circle is the inverse coupling constant $1/g$, the
gauge transformation parameter can have a periodicity $2\pi n/g$ with $n\in
\mathbb{N}$.  This is the well-known Abelian Wu-Yang monopole construction
\cite{Wu:1975es}. In our case, since the period of $\varphi$ is $2\pi$, we get
$2\pi/g $, which is the smallest value allowed $p=1/g$. The solution that
describes the monopole of charge $n$ times the minimum is $n$ times this one
$p=n/g$.

It is useful to have the expression of $B^{(\pm)}$ in Cartesian coordinates:

\begin{equation}
\label{eq:Bpm2}
B^{(\pm)} = \frac{1}{2g} \frac{[(0,0,\mp 1)\times (x^{1},x^{2},x^{3}) ]\cdot
  d\vec{x}}{r^{2}(r\pm x^{3})}\, ,
\end{equation}

\noindent
in which the singularity at $r=\mp x^{3}$ becomes evident. In this form, one
can easily change the position of the monopole from the origin to some other
point $x^{m}_{0}$ and the position of the Dirac string from the half line that
starts from the origin in the direction $-(0,0,\mp 1)$ to the half line that
starts at the monopole's position $x^{m}_{0}$ hand has the direction $s^{m}$
relative to that point:

\begin{equation}
\label{eq:Bs}
B^{(s)} 
= 
\frac{1}{2g}\left(1-\frac{s^{m}}{s}\frac{u^{m}}{u}\right)^{-1} 
\varepsilon_{mnp}\frac{s^{m}}{s}\frac{u^{n}}{u}d\frac{u^{p}}{u}\, ,
\end{equation}

\noindent
with

\begin{equation}
u^{m} \equiv x^{m}-x^{m}_{0}\, ,
\hspace{.5cm}
u^{2}\equiv u^{m}u^{m}\, ,
\hspace{.5cm}
s^{2} \equiv s^{m}s^{m}\, .
\end{equation}


\subsection{From the Dirac monopole to the Wu-Yang $\mathrm{SU}(2)$
  monopole}

Let us consider the Abelian $B^{(+)}$ solution in Eq.~(\ref{eq:Bpm})
and let us embed it in $\mathrm{SU}(2)$ as the 3rd component of the
gauge field

\begin{equation}
A^{(+)}\equiv 2 B^{(+)}T_{3}\, , 
\hspace{1cm}
F(A^{(+)}) = 2 F(B)T_{3}\, .
\end{equation}

\noindent
The $\mathrm{SU}(2)$ gauge transformation (which is evidently singular
along the negative $z$ axis and makes the whole Dirac string
singularity, but the endpoint at the coordinate origin, disappear)

\begin{equation}
U^{(+)} \equiv \frac{1}{\sqrt{2(1+\frac{z}{r})}}\left[1+\frac{z}{r}+2\left(\frac{x}{r}T_{2}-\frac{y}{r}T_{1}\right)\right]\, ,  
\end{equation}

\noindent
relates the gauged field $A^{(+)}$ to 

\begin{equation}
\label{eq:WYmonopole}
A = \frac{1}{g} \varepsilon^{a}{}_{mn}dx^{m}\frac{x^{n}}{r^{2}} T_{a}\, ,
\hspace{1cm}
A^{(+)} = U^{(+)}A (U^{(+)})^{-1} +\frac{1}{g} dU^{(+)} (U^{(+)})^{-1}\, ,
\end{equation}

\noindent
which is the gauge field of the Wu-Yang $\mathrm{SU}(2)$ monopole. As
we have mentioned in the previous appendix, this is also the gauge
field of the Lorentzian meron Eq.~(\ref{eq:Ameron}). The gauge
transformation also relates $T_{3}$ to $\mathcal{U}$ in
Eq.~(\ref{eq:U}) and the Abelian vector

\begin{equation}
U^{(+)}U (U^{(+)})^{-1} = 2 T_{3}\, .
\end{equation}

The fact that the Lorentzian meron is the Wu-Yang monopole, which is
related by a gauge transformation to the Dirac monopole makes the
relation Eq.~(\ref{eq:emtensorsrelated}) trivial.

This construction can be generalized to more general positions of the Dirac
string: if we consider embedding of the Dirac monopole solution $B^{(s)}$ in
Eq.~(\ref{eq:Bs}) into $\mathrm{SU}(2)$ 

\begin{equation}
\label{eq:Diracembedded}
A^{(s)} \equiv -2B^{(s)} \frac{s^{m}}{s} \delta_{m}{}^{a} T_{a}\, ,   
\end{equation}

\noindent
it is easy to see that the gauge transformation 

\begin{equation}
\label{eq:Us}
U^{(s)}
\equiv
\frac{1}{\sqrt{2\left(1-\frac{s^{m}}{s}\frac{u^{m}}{u}\right)}}
\left[1-\frac{s^{m}}{s}\frac{u^{m}}{u}
-2\varepsilon_{mn}{}^{a}\frac{s^{m}}{s}\frac{u^{n}}{u}T_{a}
\right]\, ,    
\end{equation}

\noindent
relates it to the same Wu-Yang monopole field Eq.~(\ref{eq:WYmonopole})

\begin{equation}
A^{(s)}
= 
U^{(s)}A (U^{(s)})^{-1} +\frac{1}{g} dU^{(s)} (U^{(s)})^{-1}\, .
\end{equation}


\section{The $\mathrm{SU}(2)$ Skyrme model}
\label{app-skyrme}

In this appendix we are going to show that the Lorentzian meron
(Wu-Yang monopole) is also associated to a solution of the equations
of motion of the $\mathrm{SU}(2)$ Skyrme model \cite{Skyrme:1962vh}
written in the form \cite{Canfora:2013xja}

\begin{equation}
S_{\rm Skyrme}
=
-\tfrac{1}{2}\int d^{4}x
\left\{
\tfrac{1}{2} R_{\mu}R^{\mu} +\frac{\lambda}{16}S_{\mu\nu}S^{\mu\nu}
\right\}\, ,
\end{equation}

\noindent
where

\begin{equation}
R_{\mu}
\equiv
V^{-1}\partial_{\mu} V\, ,
\hspace{1cm}   
S_{\mu\nu}
\equiv
[R_{\mu},R_{\nu}]\, ,
\hspace{1cm}
V(x) \in \mathrm{SU}(2)\, .
\end{equation}

The equations of motion are 

\begin{equation}
\partial_{\mu}R^{\mu} +\frac{\lambda}{4}\partial_{\mu}[R_{\nu},F^{\mu\nu}]
=
0\, .  
\end{equation}

If we take $V=U^{-1}$ ($U$ given by Eq.~(\ref{eq:U})), then we can write $R=2g
A$ where $A$ is Lorentzian meron's gauge field Eq.~(\ref{eq:Ameron}) and

\begin{equation}
  \begin{array}{rcl}
\partial_{\mu}R^{i\, \mu} 
& = & 
-2g \partial_{m}A^{i}{}_{m}=0\, ,  
\\
& & \\
\partial_{\mu}[R_{\nu},F^{\mu\nu}]^{i} 
& \sim &
\partial_{m} {\displaystyle \left(\frac{A^{i}{}_{m}}{r^{2}}\right)} =0\, .
\end{array}
\end{equation}


\section{Higher-charge Lorentzian merons and Wu-Yang monopoles}
\label{app-higherchargeWY}

The construction of a Lorentzian meron can be generalized by using a
generalization of the unit outward-pointing vector $x^{m}/r$ denoted
by $\xi^{m}$ and defined by \cite{Bais:1976fr}

\begin{equation}
(\xi^{m} )
\equiv
\frac{1}{r}
\left(
\frac{\Im\mathfrak{m}(x^{2}+ix^{1})^{n}}{\rho^{n-1}}, 
\frac{\Re\mathfrak{e}(x^{2}+ix^{1})^{n}}{\rho^{n-1}},
x^{3}
\right)\, , 
\hspace{1cm}
\rho^{2} \equiv (x^{1})^{2} +(x^{2})^{2}\, ,
\end{equation}

\noindent
or, in spherical coordinates,

\begin{equation}
(\xi^{m} )
\equiv
\left(
\sin{\theta} \sin{n\varphi}, 
\sin{\theta} \cos{n\varphi}, 
\cos{\theta}
\right)\, , 
\end{equation}

\noindent
and which reduces to $x^{m}/r$  for $n=1$. The essential properties of
$\xi^{m}$ are

\begin{eqnarray}
d\xi^{m} \wedge d\xi^{n} 
& = & 
-n\varepsilon_{mnp} \xi^{p}d\Omega^{2}\, ,
\\
& & \nonumber \\
- \tfrac{1}{2} \varepsilon_{mnp} \xi^{m} d\xi^{n} \wedge d\xi^{p} 
& = & 
nd\Omega^{2} = \star_{(3)} d\frac{n}{r}\, ,   
\end{eqnarray}

The generalization of the meron solution is constructed in terms of the
generalization $\mathrm{SU}(2)$ matrix in Eq.~(\ref{eq:U})

\begin{equation}
\label{eq:Un}
U_{(n)} \equiv 2\xi^{m}\delta_{m}^{a} T_{a}\, ,     
\hspace{1cm}
U_{(n)}^{\dagger}=U_{(n)}^{-1}= -U_{(n)}\, ,
\end{equation}

\noindent
and takes the form 

\begin{equation}
\label{eq:Ameronn}
A
\equiv 
\frac{1}{2g}dU_{(n)} U_{(n)}^{-1}\, . 
\end{equation}

\noindent
The field strength is given by 

\begin{equation}
\label{eq:F(A)n}
F(A_{(n)}) = \tfrac{1}{2}dA = g[A,A] = \star_{(3)} d \frac{n}{2gr}\, U_{(n)}\, ,
\end{equation}

\noindent
and can be related to that of a Dirac monopole of charge $p=n/g$ 

\begin{equation}
F(B_{(n)})=\star_{(3)} d \frac{n}{2gr}\, ,  
\,\,\,\,
F(A_{(n)}) = F(B_{(n)})\, U_{(n)}\, ,
\end{equation}

\noindent
which is given by the expressions studied at the beginning. The
energy-momentum tensor of $A$ is also equal to that of the Abelian monopole of
charge $n/g$ $B$. These fields can also be related to the embedding of the
charge $n/g$ Dirac monopole into $\mathrm{SU}(2)$ with a generalization of the gauge
transformation Eq.~(\ref{eq:Us})

\begin{equation}
\label{eq:Usn}
U_{(n)}^{(s)}
\equiv
\frac{1}{\sqrt{2\left(1-\frac{s^{m}}{s}\xi^{m}\right)}}
\left[1-\frac{s^{m}}{s}\xi^{m}
-2\varepsilon_{mn}{}^{a}\frac{s^{m}}{s}\xi^{n} T_{a}
\right]\, ,    
\end{equation}

\noindent
relates it to the meron gauge field:

\begin{equation}
U_{(n)}^{(s)}U_{(n)} (U_{(n)}^{(s)})^{-1} = -2\frac{s^{m}}{s} \delta_{m}{}^{a} T_{a} \, ,
\hspace{1cm}
U_{(n)}^{(s)}A_{(n)} (U_{(n)}^{(s)})^{-1} +\frac{1}{g} dU_{(n)}^{(s)} (U_{(n)}^{(s)})^{-1}
=
n B_{(n)}^{(s)} 2\frac{s^{m}}{s} \delta_{m}{}^{a} T_{a}\, .
\end{equation}

To check that this gauge field solves the Yang-Mills equations of motion we
first stress that, with the above connection, $U_{(n)}$ is a covariantly-constant
adjoint field. Then, auxiliary the adjoint Higgs field

\begin{equation}
\label{eq:Phimeronn}
\Phi_{(n)} \equiv \left(-\frac{\mu}{2g} +\frac{n}{2gr}\right)\, U_{(n)}\, ,  
\end{equation}

\noindent
satisfies

\begin{equation}
D\Phi_{(n)} = d\frac{n}{2gr}\,  U_{(n)}\, ,  
\end{equation}

\noindent
and the pair $A_{(n)},\Phi_{(n)}$ satisfies the Bogomol'nyi equations (\ref{eq:Beqs}) and,
as a consequence the equations of motion of the Yang-Mills-Higgs system. The
last equation implies that $\Phi_{(n)}$ and $D\Phi_{(n)}$ commute so the Higgs current
vanishes and $A_{(n)}$ also solves the sourceless Yang-Mills equations.


\end{document}